\pdfoutput=1
\PassOptionsToPackage{table}{xcolor} 
\documentclass[11pt]{article}

\usepackage[final]{acl}
\usepackage{listings}
\usepackage{xcolor}

\definecolor{codegreen}{rgb}{0,0.6,0}
\definecolor{codegray}{rgb}{0.5,0.5,0.5}
\definecolor{codepurple}{rgb}{0.58,0,0.82}
\definecolor{backcolour}{rgb}{1,1,1}

\lstdefinestyle{mystyle}{
    backgroundcolor=\color{backcolour},   
    commentstyle=\color{codegreen},
    keywordstyle=\color{magenta},
    numberstyle=\tiny\color{codegray},
    stringstyle=\color{codepurple},
    basicstyle=\ttfamily\footnotesize,
    breakatwhitespace=false,         
    breaklines=true,                 
    captionpos=b,                    
    keepspaces=true,                 
    numbers=left,                    
    numbersep=5pt,                  
    showspaces=false,                
    showstringspaces=false,
    showtabs=false,                  
    tabsize=2
}
\usepackage{times}
\usepackage{latexsym}

\usepackage[T1]{fontenc}

\usepackage[utf8]{inputenc}

\usepackage{microtype}

\usepackage{inconsolata}

\usepackage{graphicx}

\usepackage{inconsolata}
\usepackage{amsmath}
\usepackage{algorithm}
\usepackage{subcaption}
\usepackage{algpseudocode}
\usepackage{amsmath}
\usepackage{amsfonts}
\usepackage{float}
\usepackage{soul}

\usepackage{amsmath}
\usepackage{array}
\usepackage{tcolorbox} 
\usepackage{multirow}
\usepackage{titlesec}
\usepackage{times}
\usepackage{latexsym}
\usepackage{graphicx}
\usepackage{amsmath}
\usepackage{adjustbox}
\usepackage{graphicx}
\usepackage{booktabs}

\usepackage{xcolor} 
\usepackage{colortbl} 
\definecolor{lightlime}{RGB}{179, 255, 179} 

\title{SRLCG: Self-Rectified Large-Scale Code Generation with Multidimensional Chain-of-Thought and Dynamic Backtracking}

\author{\textbf{Hongru Ma$^1$, Yanjie Liang$^2$, Jiasheng Si$^3$, Weiyu Zhang$^3$, Hongjiao Guan$^3$} \\
\textbf{Chaoqun Zheng$^3$, Bing Xu$^4$, Wenpeng Lu$^{3,\dagger}$}
\\ \small $^1$Beihang University, China
\\ \small $^2$Shandong University, China
\\ \small $^3$Qilu University of Technology (Shandong Academy of Sciences), China
\\ \small $^4$Harbin Institute of Technology, China
}

\begin{document}
\maketitle

\def\thefootnote{$\dagger$}\footnotetext{Corresponding Author. Email: wenpeng.lu@qlu.edu.cn}
\def\thefootnote{$\ddagger$}
\begin{abstract}
Large language models (LLMs) have revolutionized code generation, significantly enhancing developer productivity. However, for a vast number of users with minimal coding knowledge, LLMs provide little support, as they primarily generate isolated code snippets rather than complete, large-scale project code. Without coding expertise, these users struggle to interpret, modify, and iteratively refine the outputs of LLMs, making it impossible to assemble a complete project. To address this issue, we propose Self-Rectified Large-Scale Code Generator (SRLCG), a framework that generates complete multi-file project code from a single prompt. SRLCG employs a novel multidimensional chain-of-thought (CoT) and self-rectification to guide LLMs in generating correct and robust code files, then integrates them into a complete and coherent project using our proposed dynamic backtracking algorithm. 
Experimental results show that SRLCG generates code \textbf{15×} longer than DeepSeek-V3, \textbf{16×} longer than GPT-4, and at least \textbf{10×} longer than other leading CoT-based baselines. Furthermore, they confirm its improved correctness, robustness, and performance compared to baselines in large-scale code generation.\footnote{Our code and dataset are available at \url{https://anonymous.4open.science/r/SRLCG-FB01}}
\end{abstract}

\section{Introduction}
The recent emergence and rise of large language models (LLMs)~\cite{wang2024historydevelopmentprincipleslarge,zhao2024surveylargelanguagemodels}, such as GPT-4~\cite{openai2024gpt4technicalreport} and DeepSeek-V3~\cite{deepseekai2024deepseekv3technicalreport}, has significantly facilitated developers in code generation, making the process more efficient and accessible.
However, for the vast majority of users without programming knowledge, the presence of LLMs has done little to alleviate their difficulties. Simply prompting LLMs to generate code is far from sufficient, as the vast majority of users, lacking the necessary programming expertise, are unable to effectively interpret, modify, and iteratively refine the generated outputs. Worse still, due to inherent limitations in LLMs, such as inconsistencies in reasoning~\cite{xie2024Calibrating,saxena2024evaluatingconsistencyreasoningcapabilities} and susceptibility to generating syntactically correct but logically flawed code~\cite{wang2024largelanguagemodelsfail,tian2025codehaluinvestigatingcodehallucinations}, the accuracy of the output remains fundamentally unreliable. This renders it nearly impossible for non-expert users to produce robust and complete code on their own.

\begin{figure}[t]
  \centering\includegraphics[height=7cm,keepaspectratio]{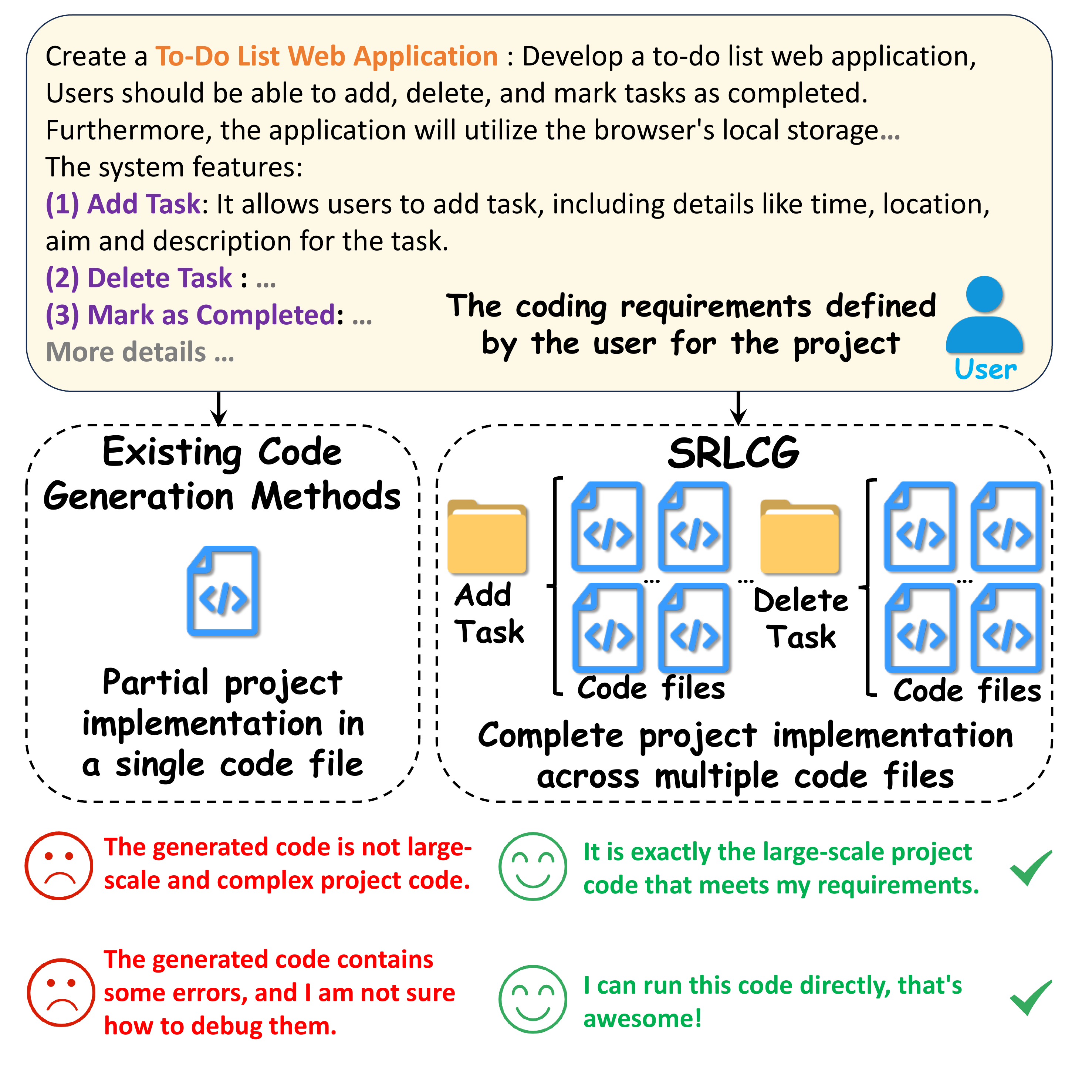} 
  \caption{Comparison of SRLCG and existing methods. Existing methods struggle with multi-file project generation while ensuring correctness and robustness, whereas SRLCG directly generates correct and robust large-scale project code.}
  \label{fig:1}
\end{figure}

Numerous studies have investigated leveraging Chain-of-Thought (CoT)~\cite{Wei2022Chain-of-thoughtprompting} prompting to enhance the code generation capabilities of LLMs, aiming to effectively elicit their full potential and address the challenges inherent in directly generating code with LLMs. CoT-based approaches contribute to improvements in various aspects, such as \textit{structured reasoning}~\cite{Zheng2023Outline,Weir2024Learning,Li2025Structured}, where multi-step reasoning guides LLMs to produce logically coherent code; \textit{problem decomposition}~\cite{jiang2024self,chen2024divideandconquer,Yen2024CoLadder}, where complex programming tasks are broken down into smaller, more manageable subproblems to improve solution quality; and \textit{contextual understanding}~\cite{zhao2024enhancing,Du2024Class}, where enriched prompts provide LLMs with more comprehensive contextual information, enhancing the accuracy of function synthesis and algorithmic implementation.

Figure~\ref{fig:1} illustrates that, although significant progress has been made in code generation, most of existing code generation methods still face two critical gaps, hindering their effective application in real-world scenarios. \textbf{Gap 1:} \textit{They are fundamentally incapable of directly generating large-scale, complex project code, which significantly hampers their practical applicability and fails to meet the demands of a broad range of users.} 
In practice, users, particularly those with little to no coding knowledge, may seek to leverage LLMs to generate complex frameworks, such as a chat system with modules for registration, login, data storage, etc. 
Although prior work like CoLadder~\cite{Yen2024CoLadder} has advanced ML-based code generation, it produces code of unsuitable length for real-world projects and addresses only a narrow task set without a generalizable solution for diverse programming needs.
\textbf{Gap 2:} \textit{more importantly, existing code generation methods often inherently neglect the overall integrity and correctness of the project code, failing to rectify the critical interactions between modules and the potential errors within them, thereby severely compromising the performance and reliability of the project code.} 
Although prior studies, such as~\cite{le2023codechain},~\cite{huang2024codecottacklingcodesyntax}, and~\cite{chen2024divideandconquer}, have focused on self-checking in code generation, they primarily focus on post-generation error detection while lacking rigorous verification of CoT rationales during the code generation process. 
As a result, this limitation compromises correctness in large-scale code generation, leading to persistent, hard-to-trace bugs that demand excessive debugging, even from experienced developers, reducing overall productivity.

In this paper, we aim to bridge the aforementioned critical gaps and address a challenge that, to the best of our knowledge, remains unresolved in prior work: \ul{\textit{How can we develop a generalizable approach for generating robust, correct large-scale project code adaptable to diverse tasks?}} To achieve these objectives, we propose a novel unified framework, namely, \textbf{Self-Rectified Large-Scale Code Generator (SRLCG)}, which enables users to generate a complete multi-file project with a single prompt, eliminating the need for coding expertise. SRLCG consists of two key modules: (1) MdCoT-DB, a \textbf{m}ulti\textbf{d}imensional \textbf{CoT} and \textbf{d}ynamic \textbf{b}acktracking to bridge the first gap, and (2) Self-Rectification with adaptive feedback to bridge the second gap.
Specifically, MdCoT-DB adopts a top-down multidimensional decomposition, progressively breaking down the original prompt into \textit{strategic}, \textit{tactical}, and \textit{operational} dimensions. This process generates atomic-level sub-function code files, dynamically integrated into a complete project via a dynamic backtracking algorithm. Meanwhile, Self-Rectification applies adaptive feedback for attenuation-based verification at each dimension, regenerating rationales if self-checking fails. Experimental results demonstrate that SRLCG surpasses all baselines in large-scale multi-project code generation, achieving superior completeness, correctness, usability, and robustness. Our key contributions are as follows.

\begin{figure*}[t]
  \centering
  \includegraphics[height=8cm,keepaspectratio]{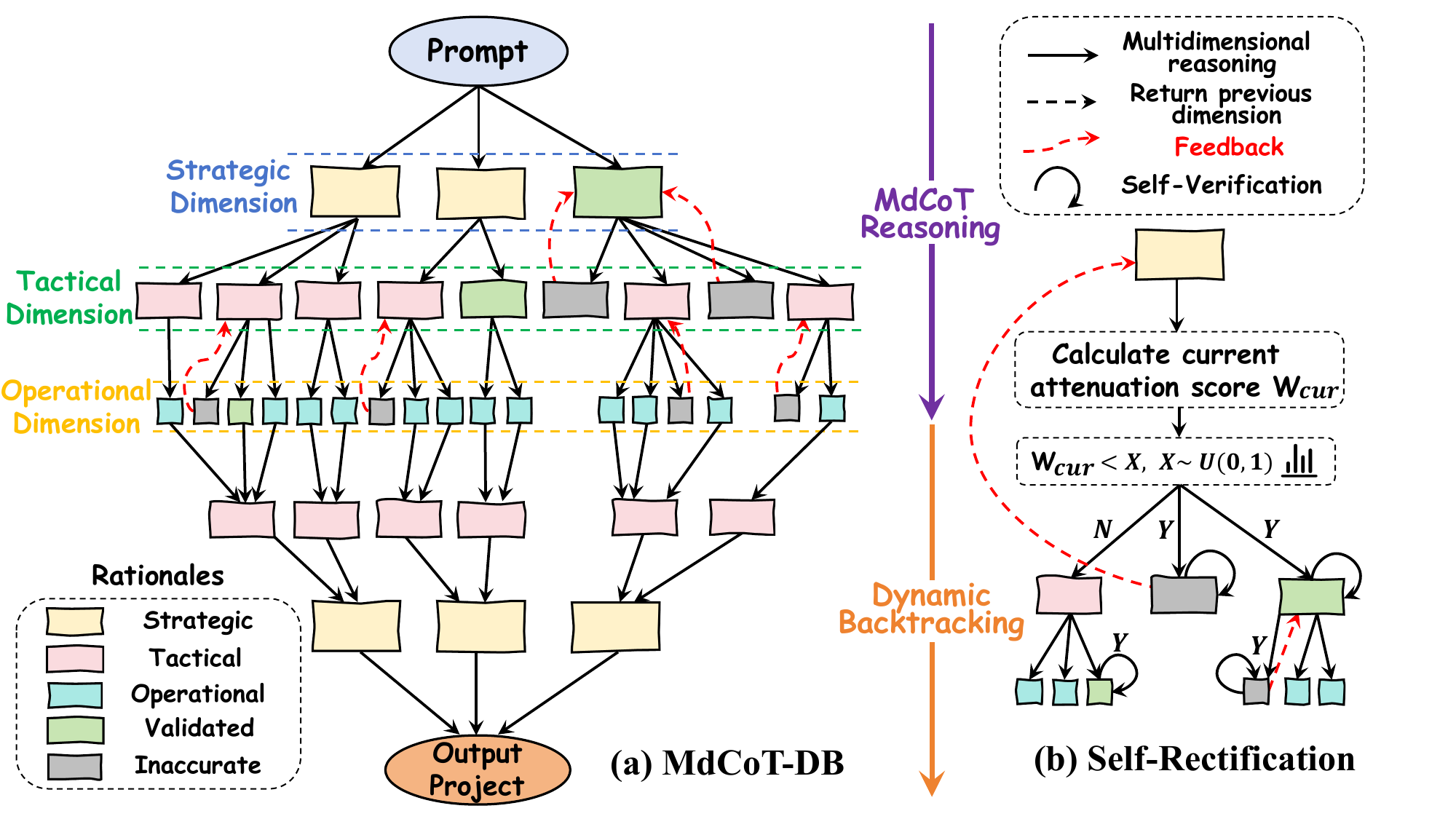} 
  \caption{The SRLCG framework consists of two modules: (a) MdCoT-DB, which decomposes user prompts into sub-functions and then integrates them via dynamic backtracking, and (b) Self-Rectification applies to each MdCoT dimension, performing verification and rectification to ensure the correctness and robustness of the generated code.}
  \label{fig:framework}
\end{figure*}

\begin{itemize}
    \setlength{\itemsep}{0pt}
    \item We propose SRLCG, a novel unified framework for large-scale multi-project code generation, ensuring correctness and robustness while addressing unresolved large-scale code generation challenges in prior work.
    \item We introduce a multidimensional CoT reasoning method to enhance LLMs' generative potential, integrating dynamic backtracking for adaptive rationale integration and progressive attenuation-based verification to balance inference time with correctness and robustness.
    \item We conduct extensive experiments across diverse tasks, showing that SRLCG generates 15× longer code than DeepSeek-V3 and 16× longer than GPT-4, surpassing leading baselines while ensuring correctness and robustness in large-scale code generation.
\end{itemize}

\section{Related Work}
\subsection{CoT-Based Code Generation}
Chain-of-Thought (CoT) reasoning has been explored in code generation to enhance logical correctness and error handling. By structuring intermediate reasoning steps, recent studies~\cite{Yen2024CoLadder,Weir2024Learning,wang2024largelanguagemodelsfail,Li2025Structured,tian2025codehaluinvestigatingcodehallucinations} aim to improve model interpretability and solution accuracy.
Several works integrate CoT into code generation. 
For example,
~\citet{le2023codechain} proposed CodeChain which employs multi-step reasoning to plan and verify code structures before generation.
~\citet{Yang2024COTTON} proposed COTTON, which leverages high-quality CoT reasoning to enhance neural code generation in lightweight language models, thereby improving their ability to generate structured and accurate code.
~\citet{Li2025Structured} proposed SCoT, which integrates program structure into CoT reasoning, enabling models to generate code through structured intermediate steps.
However, these methods remain unsuitable for large-scale project code generation. They focus on isolated problems, confined to localized code generation without system-level reasoning or architectural synthesis~\cite{stechly2024chainthoughtlessnessanalysiscot}. Moreover, their single-dimensional CoT lacks the depth to handle diverse requirements and cross-file dependencies, making them brittle and impractical for real-world, multi-file development.

\subsection{Self-Checking in Code Generation}
Self-checking mechanisms, both CoT-based and non-CoT-based, have been explored in code generation to enhance correctness and robustness. 
Non-CoT-based methods~\cite{zhang2023selfeditfaultawarecodeeditor,jiang2024ledex,chen2024teaching,wen2025fixingfunctionlevelcodegeneration,adnan2025largelanguagemodelguided} typically rely on execution feedback or static analysis to detect and correct errors. For example,~\citet{chen2024teaching} proposed Self-Debug to iteratively refine generated code by running test cases and fixing detected issues.
CoT-based approaches~\cite{le2023codechain,ling2023deductive,huang2024codecottacklingcodesyntax,chen2024divideandconquer}, on the other hand, incorporate intermediate reasoning steps to guide self-verification. For instance,
~\citet{huang2024codecottacklingcodesyntax} introduced CodeCoT to leverage CoT reasoning to identify and correct syntax errors before finalizing code generation.
However, both approaches remain inadequate for large-scale code generation. Their reliance on sequential verification lacks an effective prioritization mechanism, making rational assessment impractical as complexity increases. Moreover, their emphasis on syntax errors overlooks higher-level logical consistency. Consequently, they fail to ensure structural coherence across complex, multi-file projects.
\section{Methodology}
\subsection{Task Definition}
We define large-scale code generation as taking a given prompt \(\mathcal{P}\) as input and generating an entire project \(\mathcal{A}\) as output, including multiple directories and code files. Specifically, \(\mathcal{A}\) represented as \( \mathcal{A} = \{ \mathcal{M}_1, \mathcal{M}_2, \dots, \mathcal{M}_n \} \), where each \( \mathcal{M}_i \) corresponds to a code module that contributes to the overall solution. Each module \( \mathcal{M}_i \) is further decomposed into a set of functions, expressed as \( \mathcal{M}_i = \{ \mathcal{F}_1, \mathcal{F}_2, \dots, \mathcal{F}_m \} \), where each \( \mathcal{F}_j \) represents a function that contributes to the overall module. The generation process of each function \( \mathcal{F}_j \) is outlined as follows:
\begin{equation}
\mathcal{F}_j = \arg\max_{\hat{y}} \prod_{k \in D_l} P_{\text{LLM}_{\theta}}(\hat{y} \mid \mathcal{T}_{\text{MdCoT}_{k}}, \mathcal{P}),
\end{equation}
where \( \mathcal{T}_{\text{MdCoT}_{k}} \) denotes the intermediate MdCoT\footnote{Details are provided in Section \ref{sec:MdCoT-DB}.} prompt associated with each dimension, and \( \hat{y} \) represents the set of all possible outcomes of \( \mathcal{F}_j \) through all the dimensions \( D_l \) from MdCoT, with reference to the $\text{LLM}_{\theta}$.

After that, the goal \(\mathcal{A}\) is defined in the following:
\begin{equation}
    \mathcal{A} = \bigcup_{i=1}^{n} \mathcal{M}_i, \quad \text{with} \quad \mathcal{M}_i = \bigcup_{j=1}^{m} \mathcal{F}_j,
\end{equation}
where each \( \mathcal{F}_j \) contributes to the construction of \( \mathcal{M}_i \), and each \( \mathcal{M}_i \) in turn contributes to the formation of \( \mathcal{A} \), all of which are refined and integrated through dynamic backtracking.

\subsection{Overview}
As shown in Figure~\ref{fig:framework}, SRLCG consists of two key modules: (1) \textit{Multidimensional CoT and Dynamic Backtracking} (MdCoT-DB), which employs Multidimensional CoT to decompose the original prompt into three distinct levels of granularity (\textbf{strategic}, \textbf{tactical}, and \textbf{operational}), ultimately generating multiple code files. These files are then synthesized into a complete project through dynamic backtracking. (2) \textit{Self-Rectification}, which operates directly on MdCoT, using adaptive feedback to dynamically assign weights and determine whether the rationale of MdCoT needs verification. If verification is required and fails, the reasoning process rolls back to the previous dimension to regenerate the rationale.

\subsection{MdCoT-DB}
\label{sec:MdCoT-DB}
In the MdCoT-DB module, we draw inspiration from the divide-and-conquer paradigm to decompose large-scale code generation into two distinct stages, facilitating a more systematic and efficient approach. As shown in Figure~\hyperref[fig:framework]{\ref{fig:framework}(a)}, we specialize the divide-and-conquer paradigm as follows: \textbf{Divide}: A multidimensional CoT (\ref{sub:MdCoT}) elicits the LLM’s ability to reason about the project’s complexity across multiple levels of abstraction and granularity; \textbf{Conquer}: A dynamic backtracking algorithm (\ref{sub:DB}) progressively integrates code files from lower to higher dimensions, ultimately synthesizing them into a coherent and complete project.

\subsubsection{Multidimensional CoT}
\label{sub:MdCoT}
We define the dimensions of MdCoT, from highest to lowest, as strategic, tactical, and operational, as detailed below.

\textbf{Dimension 1: Strategic Dimension.} The strategic dimension represents the highest-level phase of task decomposition and architectural design. The process begins with a thorough analysis of the input prompt \(\mathcal{P}\), which encapsulates the general requirements, objectives, and constraints of the task. Based on this analysis, the task is systematically decomposed into several rationales \(\mathcal{R}_{\mathcal{M}_i}\) of key modules \(\mathcal{M}_i\), each responsible for handling a distinct aspect of the task:
\begin{equation}
\mathcal{R}_{\mathcal{M}_i} = \text{LLM}_{\theta}(\mathcal{P} \mid \mathcal{T}_{\text{MdCoT}_1}),
\end{equation}
where $\mathcal{T}_{\text{MdCoT}_1}$ represents the strategic dimension prompt.


Through the strategic dimension of MdCoT, the initial singular input prompt \(\mathcal{P}\) is systematically partitioned into \(n\) distinct rationales \(\mathcal{R}_{\mathcal{M}_i}\), each corresponding to a specific module \(\mathcal{M}_i\). 

\textbf{Dimension 2: Tactical Dimension.} The tactical dimension serves as a crucial bridge between the strategic and operational dimensions, refining the high-level structures outlined in the strategic dimension. Building upon the rationales $\mathcal{R}_{\mathcal{M}_i}$ derived from the previous dimension, this phase focuses on further decomposing each module into its constituent rationales \(\mathcal{R}_{\mathcal{F}_j}\) of sub-functions \(\mathcal{F}_j\) and defining their respective responsibilities: 
\begin{equation}
\mathcal{R}_{\mathcal{F}_j} = \text{LLM}_{\theta}(\mathcal{R}_{\mathcal{M}_i} \mid \mathcal{T}_{\text{MdCoT}_2}),
\end{equation}
where $\mathcal{T}_{\text{MdCoT}_2}$ represents the tactical dimension prompt.

Through the tactical dimension of MdCoT, each rationale $\mathcal{R}_{\mathcal{M}_i}$ undergoes further decomposition into rationales $\mathcal{R}_{\mathcal{F}_j}$ of sub-functions \(\mathcal{F}_j\), contributing to the construction of module $\mathcal{M}_i$. 

\textbf{Dimension 3: Operational Dimension.} The Operational Dimension represents the final stage of reasoning within MdCoT, where the sub-function rationales $\mathcal{R}_{\mathcal{F}}$ derived from the tactical dimension are directly converted into executable code:
\begin{equation}
\mathcal{F}_j = \arg\max_{\hat{y}} P_{\text{LLM}_{\theta}}(\hat{y} \mid \mathcal{T}_{\text{MdCoT}_{3}}, \mathcal{R}_{\mathcal{F}_j}),  
\end{equation}
where $\mathcal{T}_{\text{MdCoT}_3}$ represents the operational dimension prompt.

Through the operational dimension of MdCoT, each rationale $\mathcal{R}_{\mathcal{F}_j}$ is leveraged to generate the corresponding sub-function $\mathcal{F}_j$, thereby ensuring precise and coherent implementation. 

\subsubsection{Dynamic Backtracking Algorithm}
\label{sub:DB}

We propose a new dynamic backtracking algorithm to address the challenge of correctly synthesizing a complete project from discrete sub-functions generated through MdCoT. Our approach, detailed in Algorithm \ref{alg:hierarchical-dynamic-backtracking}, iteratively refines function-level rationales and propagates corrections across modules, ultimately synthesizing them into a complete and coherent project.
At each iteration, we first construct module-level representations by merging function-level rationales and pushing them to a stack (lines 4–6). If a conflict is detected within a module, we identify a conflict set, which represents issues such as dependency mismatches, logical inconsistencies, and return type discrepancies, and determine the affected subset requiring revision (lines 7–9). To resolve inconsistencies, we invoke the backbone LLM, which refines the affected functions while preserving structural dependencies (line 10). The revised module is then integrated into the global reasoning process (line 12).
Similarly, we extend this mechanism to the project level, identifying and correcting inconsistencies across modules (lines 14–20). The process continues until convergence criteria are met, leading to a finalized, conflict-free project representation.
\subsection{Self-Rectification}

Due to the inherent instability of LLMs~\cite{zhou2024relying,Huang_2025}, self-rectification is required to operate on the rationales generated at each dimension to enhance reliability during code generation. However, unlike previous work, our focus on large-scale code generation verification necessitates a method that balances inference efficiency and code correctness.
Considering the complexity of large-scale code generation, which involves multiple inference steps, and the empirical observation that models often generate correct results after self-rectification, we propose an adaptive feedback mechanism for self-rectification. In this mechanism, the rectification weight \(\mathcal{W}^d\) represents the probability of verification at each reasoning step within dimension $d$ in MdCoT, progressively attenuating over successive iterations.
Formally, the progressive attenuation is defined below:
\begin{equation}
\begin{split}
\mathcal{W}^d_{\text{new}} = \\ \max\Bigl(&\mathcal{W}^d_{\text{min}}, \mathcal{W}^d_{\text{current}} \times \left(1 - \alpha \times f \right)^{\beta \times I^d}\Bigr).
\end{split}
\end{equation}
Here, in dimension $d$ of MdCoT, \(\mathcal{W}^d_{\text{current}}\) is the current weight in reasoning, while \(\mathcal{W}^d_{\text{new}}\) results from progressive attenuation. \(\alpha\) is the base attenuation coefficient, \(f\) the self-rectification frequency, \(\beta\) the significance adjustment, \(I^d\) the feedback impact score of dimensoin $d$, and \(\mathcal{W}^d_{\text{min}}\) the minimum weight threshold of dimension $d$.

\begin{algorithm}[t]
\small
\caption{Dynamic Backtracking}
\label{alg:hierarchical-dynamic-backtracking}
\begin{algorithmic}[1]
\Require Function-level rationales $\mathcal{R} = \{ \mathcal{F}_1, \mathcal{F}_2, ..., \mathcal{F}_m \}$  
\Ensure Consistent project-level reasoning $\mathcal{A}^*$  

\State Initialize function-level stack $S_F = \emptyset$, module-level stack $S_M = \emptyset$, step index $t = 0$  

\Repeat
    \State Initialize module set $\mathcal{M} = \emptyset$  
    \For{each module $\mathcal{M}_i$ in project $\mathcal{A}$}
        \State Merge $\mathcal{M}_i = \{ \mathcal{F}_1, \mathcal{F}_2, ..., \mathcal{F}_m \}$  
        \State Push $\mathcal{M}_i$ to $S_F$  
        
        \If{conflict detected in $\mathcal{M}_i$}  
            \State Identify conflict set $\mathcal{F}_\text{conf}$  
             
            \State Determine affected subset $\mathcal{F}_\text{aff}$  
           
            \State Invoke LLM: $\mathcal{F}_\text{aff}^* = \text{LLM}(\mathcal{F}_\text{conf}, \mathcal{F}_\text{aff})$  

        \EndIf  
        
        \State Merge updated $\mathcal{M}_i$ into $\mathcal{M}$  
    \EndFor  
    \State Push $\mathcal{M}$ to $S_M$  

    \If{conflict detected in project $\mathcal{A}$}  
        \State Identify conflict set $\mathcal{M}_\text{conf}$  
        \State Determine affected subset $\mathcal{M}_\text{aff}$  
        \State Invoke LLM: $\mathcal{M}_\text{aff}^* = \text{LLM}(\mathcal{M}_\text{conf}, \mathcal{M}_\text{aff})$  

    \EndIf  

    \State Merge updated $\mathcal{M}$ into global project reasoning $\mathcal{A}^*$, increment $t$  
\Until{convergence or max iterations reached}  

\State \Return Completed Project $\mathcal{A}^*$  

\end{algorithmic}
\end{algorithm}

As shown in Figure~\hyperref[fig:framework]{\ref{fig:framework}(b)}, based on adaptive feedback, self-rectification refines the reasoning process by allowing each dimension of MdCoT to revisit and improve its own intermediate rationales. Specifically, for each dimension \( d \), the weight \( \mathcal{W}^d \) is compared against a randomly sampled value from a uniform distribution \( \mathbf{U}(0,1) \). If \( \mathcal{W}^d \) is smaller, self-verification is triggered, and if the verification fails, a feedback loop is introduced to further refine the rationale.

\section{Experiments}
\subsection{Experimental Setup}

\begin{table*}[ht]
\centering
\small 
\begin{adjustbox}{width=1\textwidth, center} 
\begin{tabular}{c|l|c|c|c|c|c|c}
\toprule
\textbf{Backbone} & \textbf{Method} & \textbf{Game} & \textbf{Web} & \textbf{AI/ML} & \textbf{DataBase} & \textbf{Mobile} & \textbf{Average} \\
\midrule
\multirow{9}{*}{DeepSeek-V3} 
 & Vanilla LLM & 3,547 & 3,218 & 3,329 & 3,456 & 3,631 & 3,436 \\
 & CoT-pmt & 4,219\textsubscript{\textcolor{red}{$\uparrow$$672$}} & 4,032\textsubscript{\textcolor{red}{$\uparrow$$814$}} & 4,307\textsubscript{\textcolor{red}{$\uparrow$$978$}} & 3,845\textsubscript{\textcolor{red}{$\uparrow$$389$}} & 4,128\textsubscript{\textcolor{red}{$\uparrow$$497$}} & 4,106\textsubscript{\textcolor{red}{$\uparrow$$670$}} \\
 & KQG - CoT+ & 2,200\textsubscript{\textcolor{blue}{$\downarrow$$1347$}} & 2,056\textsubscript{\textcolor{blue}{$\downarrow$$1162$}} & 2,143\textsubscript{\textcolor{blue}{$\downarrow$$1186$}} & 2,159\textsubscript{\textcolor{blue}{$\downarrow$$1297$}} & 2,237\textsubscript{\textcolor{blue}{$\downarrow$$1394$}} & 2,159\textsubscript{\textcolor{blue}{$\downarrow$$1277$}} \\
 & SCOT(LI) & 5,223\textsubscript{\textcolor{red}{$\uparrow$$1676$}} & 5,041\textsubscript{\textcolor{red}{$\uparrow$$1823$}} & 5,117\textsubscript{\textcolor{red}{$\uparrow$$1788$}} & 4,932\textsubscript{\textcolor{red}{$\uparrow$$1476$}} & 5,059\textsubscript{\textcolor{red}{$\uparrow$$1428$}} & 5,074\textsubscript{\textcolor{red}{$\uparrow$$1638$}} \\
 & COTTON & 5,543 \textsubscript{\textcolor{red}{$\uparrow$$1996$}} & 5,321\textsubscript{\textcolor{red}{$\uparrow$$2103$}} & 5,418\textsubscript{\textcolor{red}{$\uparrow$$2089$}} & \underline{5,734}\textsubscript{\textcolor{red}{$\uparrow$$2278$}} & \underline{5,876}\textsubscript{\textcolor{red}{$\uparrow$$2245$}} & 5,578\textsubscript{\textcolor{red}{$\uparrow$$2142$}} \\
 & Self-planning & 2,059\textsubscript{\textcolor{blue}{$\downarrow$$1488$}} & 1,866\textsubscript{\textcolor{blue}{$\downarrow$$1352$}} & 1,927\textsubscript{\textcolor{blue}{$\downarrow$$1402$}} & 2,022\textsubscript{\textcolor{blue}{$\downarrow$$1434$}} & 2,107\textsubscript{\textcolor{blue}{$\downarrow$$1524$}} & 1,996\textsubscript{\textcolor{blue}{$\downarrow$$1440$}} \\
 & AceCoder & \underline{5,812}\textsubscript{\textcolor{red}{$\uparrow$$2265$}} & \underline{5,634}\textsubscript{\textcolor{red}{$\uparrow$$2416$}} & \underline{5,729}\textsubscript{\textcolor{red}{$\uparrow$$2400$}} & 5,547\textsubscript{\textcolor{red}{$\uparrow$$2091$}} & 5,668\textsubscript{\textcolor{red}{$\uparrow$$2037$}} & \underline{5,678}\textsubscript{\textcolor{red}{$\uparrow$$2242$}} \\
 & Scot(Md) & 5,034\textsubscript{\textcolor{red}{$\uparrow$$1487$}} & 4,856\textsubscript{\textcolor{red}{$\uparrow$$1638$}} & 4,923\textsubscript{\textcolor{red}{$\uparrow$$1594$}} & 4,765\textsubscript{\textcolor{red}{$\uparrow$$1309$}} & 4,887\textsubscript{\textcolor{red}{$\uparrow$$1256$}} & 4,893\textsubscript{\textcolor{red}{$\uparrow$$1457$}} \\
 & COP & 3,317\textsubscript{\textcolor{blue}{$\downarrow$$230$}} & 3,145\textsubscript{\textcolor{blue}{$\downarrow$$73$}} & 3,228\textsubscript{\textcolor{blue}{$\downarrow$$101$}} & 3,039\textsubscript{\textcolor{blue}{$\downarrow$$417$}} & 3,176\textsubscript{\textcolor{blue}{$\downarrow$$455$}} & 3,181\textsubscript{\textcolor{blue}{$\downarrow$$255$}} \\ 
 & \cellcolor{lime}\textbf{SRLCG (ours)} & \cellcolor{lime}\textbf{56,983}\textsubscript{\textbf{\textcolor{red}{$\uparrow$$53436$}}} & \cellcolor{lime}\textbf{52,831}\textsubscript{\textbf{\textcolor{red}{$\uparrow$$49613$}}} & \cellcolor{lime}\textbf{52,856}\textsubscript{\textbf{\textcolor{red}{$\uparrow$$49527$}}} & \cellcolor{lime}\textbf{55,492}\textsubscript{\textbf{\textcolor{red}{$\uparrow$$52036$}}} & \cellcolor{lime}\textbf{58,127}\textsubscript{\textbf{\textcolor{red}{$\uparrow$$54496$}}} & \cellcolor{lime}\textbf{55,257}\textsubscript{\textbf{\textcolor{red}{$\uparrow$$51821$}}} \\
\midrule
\multirow{9}{*}{GPT-4} 
 & Vanilla LLM & 3,623 & 3,456 & 3,378 & 3,512 & 3,487 & 3,491 \\
 & Cot-pmt & 4,431\textsubscript{\textcolor{red}{$\uparrow$$808$}} & 4,256\textsubscript{\textcolor{red}{$\uparrow$$800$}} & 4,345\textsubscript{\textcolor{red}{$\uparrow$$967$}} & 4,178\textsubscript{\textcolor{red}{$\uparrow$$666$}} & 4,269\textsubscript{\textcolor{red}{$\uparrow$$782$}} & 4,295\textsubscript{\textcolor{red}{$\uparrow$$804$}} \\
 & KQG - CoT+ & 2,257\textsubscript{\textcolor{blue}{$\downarrow$$1366$}} & 2,119\textsubscript{\textcolor{blue}{$\downarrow$$1337$}} & 2,154\textsubscript{\textcolor{blue}{$\downarrow$$1224$}} & 2,203\textsubscript{\textcolor{blue}{$\downarrow$$1309$}} & 2,305\textsubscript{\textcolor{blue}{$\downarrow$$1182$}} & 2,207\textsubscript{\textcolor{blue}{$\downarrow$$1284$}} \\
 & SCOT(LI) & 4,523\textsubscript{\textcolor{red}{$\uparrow$$900$}} & 4,487\textsubscript{\textcolor{red}{$\uparrow$$1031$}} & 4,476\textsubscript{\textcolor{red}{$\uparrow$$1098$}} & 4,492\textsubscript{\textcolor{red}{$\uparrow$$980$}} & 4,511\textsubscript{\textcolor{red}{$\uparrow$$1024$}} & 4,497\textsubscript{\textcolor{red}{$\downarrow$$1006$}} \\
 & COTTON & 5,123\textsubscript{\textcolor{red}{$\uparrow$$1500$}} & 5,034\textsubscript{\textcolor{red}{$\uparrow$$1578$}} & 5,067\textsubscript{\textcolor{red}{$\uparrow$$1689$}} & 4,945\textsubscript{\textcolor{red}{$\uparrow$$1433$}} & 5,045\textsubscript{\textcolor{red}{$\uparrow$$1558$}} & 5,042\textsubscript{\textcolor{red}{$\uparrow$$1551$}} \\
 & Self-planning & 5,634\textsubscript{\textcolor{red}{$\uparrow$$2011$}} & 5,456\textsubscript{\textcolor{red}{$\uparrow$$2000$}} & 5,523\textsubscript{\textcolor{red}{$\uparrow$$2145$}} & 5,378\textsubscript{\textcolor{red}{$\uparrow$$1866$}} & 5,489\textsubscript{\textcolor{red}{$\uparrow$$2002$}} & 5,496\textsubscript{\textcolor{red}{$\uparrow$$2005$}} \\
 & AceCoder & \underline{5,923}\textsubscript{\textcolor{red}{$\uparrow$$2300$}} & \underline{5,745}\textsubscript{\textcolor{red}{$\uparrow$$2289$}} & \underline{5,834}\textsubscript{\textcolor{red}{$\uparrow$$2456$}} & \underline{5,678}\textsubscript{\textcolor{red}{$\uparrow$$2166$}} & \underline{5,783}\textsubscript{\textcolor{red}{$\uparrow$$2296$}} & \underline{5,792}\textsubscript{\textcolor{red}{$\uparrow$$2301$}} \\
 & Scot(Md) & 5,345\textsubscript{\textcolor{red}{$\uparrow$$1722$}} & 5,178\textsubscript{\textcolor{red}{$\uparrow$$1722$}} & 5,256\textsubscript{\textcolor{red}{$\uparrow$$1878$}} & 5,089\textsubscript{\textcolor{red}{$\uparrow$$1577$}} & 5,190\textsubscript{\textcolor{red}{$\uparrow$$1703$}} & 5,211\textsubscript{\textcolor{red}{$\uparrow$$1720$}} \\
 & COP & 2,174\textsubscript{\textcolor{blue}{$\downarrow$$1449$}} & 2,074\textsubscript{\textcolor{blue}{$\downarrow$$1382$}} & 2,027\textsubscript{\textcolor{blue}{$\downarrow$$1351$}} & 2,107\textsubscript{\textcolor{blue}{$\downarrow$$1405$}} & 2,140\textsubscript{\textcolor{blue}{$\downarrow$$1347$}} & 2,104\textsubscript{\textcolor{blue}{$\downarrow$$1387$}} \\
 & \cellcolor{lime}\textbf{SRLCG (ours)} & \cellcolor{lime}\textbf{61,934}\textsubscript{\textbf{\textcolor{red}{$\uparrow$$58311$}}} & \cellcolor{lime}\textbf{58,901}\textsubscript{\textbf{\textcolor{red}{$\uparrow$$55445$}}} & \cellcolor{lime}\textbf{63,596}\textsubscript{\textbf{\textcolor{red}{$\uparrow$$60218$}}} & \cellcolor{lime}\textbf{60,121}\textsubscript{\textbf{\textcolor{red}{$\uparrow$$56609$}}} & \cellcolor{lime}\textbf{62,781}\textsubscript{\textbf{\textcolor{red}{$\uparrow$$59294$}}} & \cellcolor{lime}\textbf{61,466}\textsubscript{\textbf{\textcolor{red}{$\uparrow$$57975$}}} \\
\bottomrule
\end{tabular}
\end{adjustbox}
\caption{Code length comparison of SRLCG with other methods. Each data sample undergoes three trials, and the average length is reported. Code length is measured relative to the Vanilla LLM, with improvements indicated in \textcolor{red}{red} and declines in \textcolor{blue}{blue} subscripts. The best score is highlighted in \textbf{bold}, while the second-best is \underline{underlined}.}
\label{tab:model_comparison}
\end{table*}

$\bullet$ \textbf{Datasets}.
\begin{table}[t]
    \centering
    \resizebox{1\linewidth}{!}{ 
        \begin{tabular}{|l|c|}
            \hline
            \textbf{Category} & \textbf{Samples} \\
            \hline
            Game Development (Game) & 60 \\
            Web Development (Web) & 72 \\
            Artificial Intelligence / Machine Learning (AI/ML) & 100 \\
            Database Management (Database) & 60 \\
            Mobile Development (Mobile) & 60 \\
            Project Management Tool (Tool) & 48 \\
            \hline
            Total & 400 \\
            \hline
        \end{tabular}
    }
    \caption{Experimental Dataset Statistics.}
    \label{tab:sample_distribution}
\end{table}
To the best of our knowledge, no prior work has addressed large-scale code generation for complete projects, nor is there a public dataset for this task. To bridge this gap, we construct a dataset for large-scale generation, covering game development, web development, AI, mobile development, database management, and project management tools, as shown in Table~\ref{tab:sample_distribution}. The dataset includes 400 samples with task descriptions, functional requirements, and technical specifications. Further details are in Appendix~\ref{sec:Dataset}.

\noindent $\bullet$ \textbf{Evaluation Metrics}. 
Since traditional metrics such as ROUGE, which are commonly used for short-code evaluation, are inadequate for assessing large-scale code generation, we propose new evaluation metrics for large-scale code generation: (1) \textit{Code Length}, measuring the average byte size; (2) \textit{Project Completeness}, assessing completeness, correctness, usability, and robustness on a 0-to-100 scale via LLM assessment; (3) \textit{Human Evaluation}, incorporating project satisfaction, code quality improvement, and task completion rate, all evaluated on a 0-to-100 scoring scale, along with time efficiency. More details can be found in Appendix~\ref{sec:evaluation_metrics}.

\noindent $\bullet$ \textbf{Baselines}. We compare our method with several leading CoT prompting techniques and code generation approaches. Specifically, we include Vanilla LM, COTTON~\cite{Yang2024COTTON}, Self-planning~\cite{jiang2024self}, AceCoder~\cite{Li2024AceCoder}, Chain-of-thought prompting~\cite{Wei2022Chain-of-thoughtprompting}, KQG-CoT+~\cite{liang2023prompting}, Scot(Md)~\cite{Md2024Scot}, COP~\cite{Xu2024COP}, and SCOT(LI)~\cite{Li2025Structured}. More details can be found in Appendix~\ref{sec:baselines}.

\noindent $\bullet$ \textbf{Backbone Models}.
We apply our SRLCG method to different LLMs: GPT-4 and DeepSeek-V3. GPT-4 is known for its strong reasoning and code generation capabilities, while DeepSeek-V3 is an open-source model designed for long-text comprehension and structured reasoning.

\begin{table*}[ht]
\centering
\begin{adjustbox}{center} 
\small
\begin{tabular}{c|l|c|c|c|c|c}
\toprule
\textbf{Backbone} & \textbf{Method} & \textbf{Completeness} & \textbf{Correctness} & \textbf{Usability} & \textbf{Robustness} & \textbf{Average} \\
\midrule
\multirow{10}{*}{DeepSeek-V3} 
 & Vanilla LLM & 75.2 & 78.6 & 72.4 & 74.3 & 75.1 \\
 & CoT-pmt & 72.5\textsubscript{\textcolor{blue}{$\downarrow$$2.7$}} & 75.3\textsubscript{\textcolor{blue}{$\downarrow$$3.3$}} & 70.8\textsubscript{\textcolor{blue}{$\downarrow$$1.6$}} & 73.1\textsubscript{\textcolor{blue}{$\downarrow$$1.2$}} & 72.9\textsubscript{\textcolor{blue}{$\downarrow$$2.2$}} \\
 & KQG-CoT+ & 71.8\textsubscript{\textcolor{blue}{$\downarrow$$3.4$}} & 78.6\textsubscript{} & 72.4\textsubscript{} & 75.9\textsubscript{\textcolor{red}{$\uparrow$$1.6$}} & 74.7\textsubscript{\textcolor{blue}{$\downarrow$$0.4$}} \\
 & SCOT(LI) & 76.2\textsubscript{\textcolor{red}{$\uparrow$$1.0$}} & 80.4\textsubscript{\textcolor{red}{$\uparrow$$1.8$}} & 78.5\textsubscript{\textcolor{red}{$\uparrow$$6.1$}} & 81.3\textsubscript{\textcolor{red}{$\uparrow$$7.0$}} & 79.1\textsubscript{\textcolor{red}{$\uparrow$$4.0$}} \\
 & COTTON & 74.9\textsubscript{\textcolor{blue}{$\downarrow$$0.3$}} & \underline{89.6}\textsubscript{\textcolor{red}{$\uparrow$$11.0$}} & 82.6\textsubscript{\textcolor{red}{$\uparrow$$10.2$}} & 84.7\textsubscript{\textcolor{red}{$\uparrow$$10.4$}} & 83.0\textsubscript{\textcolor{red}{$\uparrow$$7.9$}} \\
 & Self-planning & \underline{83.7}\textsubscript{\textcolor{red}{$\uparrow$$8.5$}} & 84.5\textsubscript{\textcolor{red}{$\uparrow$$5.9$}} & \underline{85.2}\textsubscript{\textcolor{red}{$\uparrow$$12.8$}} & \underline{88.6}\textsubscript{\textcolor{red}{$\uparrow$$14.3$}} & \underline{85.5}\textsubscript{\textcolor{red}{$\uparrow$$10.4$}} \\
 & AceCoder & 73.6\textsubscript{\textcolor{blue}{$\downarrow$$1.6$}} & 79.8\textsubscript{\textcolor{red}{$\uparrow$$1.2$}} & 76.8\textsubscript{\textcolor{red}{$\uparrow$$4.4$}} & 80.2\textsubscript{\textcolor{red}{$\uparrow$$5.9$}} & 77.6\textsubscript{\textcolor{red}{$\uparrow$$2.5$}} \\
 & Scot(Md) & 78.9\textsubscript{\textcolor{red}{$\uparrow$$3.7$}} & 81.2\textsubscript{\textcolor{red}{$\uparrow$$2.6$}} & 82.9\textsubscript{\textcolor{red}{$\uparrow$$10.5$}} & 85.4\textsubscript{\textcolor{red}{$\uparrow$$11.1$}} & 82.1\textsubscript{\textcolor{red}{$\uparrow$$7.0$}} \\
 & COP & 80.3\textsubscript{\textcolor{red}{$\uparrow$$5.1$}} & 83.5\textsubscript{\textcolor{red}{$\uparrow$$4.9$}} & 84.7\textsubscript{\textcolor{red}{$\uparrow$$12.3$}} & 87.9\textsubscript{\textcolor{red}{$\uparrow$$13.6$}} & 84.1\textsubscript{\textcolor{red}{$\uparrow$$9.0$}} \\
 & \cellcolor{lime}\textbf{SRLCG (ours)} & \cellcolor{lime}\textbf{91.7}\textsubscript{\textcolor{red}{$\uparrow$$16.5$}} & \cellcolor{lime}\textbf{91.6}\textsubscript{\textcolor{red}{$\uparrow$$13.0$}} & \cellcolor{lime}\textbf{92.3}\textsubscript{\textcolor{red}{$\uparrow$$19.9$}} & \cellcolor{lime}\textbf{93.5}\textsubscript{\textcolor{red}{$\uparrow$$19.2$}} & \cellcolor{lime}\textbf{92.3}\textsubscript{\textcolor{red}{$\uparrow$$17.2$}} \\
\midrule
\multirow{10}{*}{GPT-4} 
 & Vanilla LLM  & 76.2 & 79.3 & 73.1 & 75.2 & 76.0 \\
 & CoT-pmt & 73.6\textsubscript{\textcolor{blue}{$\downarrow$2.6}} & 76.1\textsubscript{\textcolor{blue}{$\downarrow$3.2}} & 71.4\textsubscript{\textcolor{blue}{$\downarrow$1.7}} & 74.1\textsubscript{\textcolor{blue}{$\downarrow$1.1}} & 73.3\textsubscript{\textcolor{blue}{$\downarrow$2.6}} \\
 & KQG-CoT+ & 72.1\textsubscript{\textcolor{blue}{$\downarrow$4.1}} & 79.2\textsubscript{\textcolor{blue}{$\downarrow$0.1}} & 72.9\textsubscript{\textcolor{blue}{$\downarrow$0.2}} & 76.4\textsubscript{\textcolor{red}{$\uparrow$1.2}} & 74.7\textsubscript{\textcolor{blue}{$\downarrow$1.3}} \\
 & SCOT(LI) & 77.1\textsubscript{\textcolor{red}{$\uparrow$0.9}} & 81.2\textsubscript{\textcolor{red}{$\uparrow$1.9}} & 79.6\textsubscript{\textcolor{red}{$\uparrow$6.5}} & 82.1\textsubscript{\textcolor{red}{$\uparrow$6.9}} & 80.0\textsubscript{\textcolor{red}{$\uparrow$4.0}} \\
 & COTTON & 75.4\textsubscript{\textcolor{blue}{$\downarrow$0.8}} & \underline{90.6}\textsubscript{\textcolor{red}{$\uparrow$11.3}} & 83.4\textsubscript{\textcolor{red}{$\uparrow$10.3}} & 85.6\textsubscript{\textcolor{red}{$\uparrow$10.4}} & 84.3\textsubscript{\textcolor{red}{$\uparrow$8.3}} \\
 & Self-planning & \underline{84.6}\textsubscript{\textcolor{red}{$\uparrow$8.4}} & 85.7\textsubscript{\textcolor{red}{$\uparrow$6.4}} & \underline{86.1}\textsubscript{\textcolor{red}{$\uparrow$13.0}} & \underline{89.6}\textsubscript{\textcolor{red}{$\uparrow$14.4}} & \underline{86.4}\textsubscript{\textcolor{red}{$\uparrow$10.4}} \\
 & AceCoder & 73.9\textsubscript{\textcolor{blue}{$\downarrow$2.3}} & 80.6\textsubscript{\textcolor{red}{$\uparrow$1.3}} & 77.6\textsubscript{\textcolor{red}{$\uparrow$4.5}} & 81.1\textsubscript{\textcolor{red}{$\uparrow$5.9}} & 78.3\textsubscript{\textcolor{red}{$\uparrow$2.3}} \\
 & Scot(Md) & 79.6\textsubscript{\textcolor{red}{$\uparrow$3.4}} & 82.1\textsubscript{\textcolor{red}{$\uparrow$2.8}} & 83.6\textsubscript{\textcolor{red}{$\uparrow$10.5}} & 86.1\textsubscript{\textcolor{red}{$\uparrow$10.9}} & 82.9\textsubscript{\textcolor{red}{$\uparrow$6.9}} \\
 & COP & 81.1\textsubscript{\textcolor{red}{$\uparrow$4.9}} & 84.1\textsubscript{\textcolor{red}{$\uparrow$4.8}} & 85.6\textsubscript{\textcolor{red}{$\uparrow$12.5}} & 88.6\textsubscript{\textcolor{red}{$\uparrow$13.4}} & 84.9\textsubscript{\textcolor{red}{$\uparrow$8.9}} \\
 & \cellcolor{lime}\textbf{SRLCG (ours)} & \cellcolor{lime}\textbf{92.6}\textsubscript{\textcolor{red}{$\uparrow$16.4}} & \cellcolor{lime}\textbf{92.4}\textsubscript{\textcolor{red}{$\uparrow$13.1}} & \cellcolor{lime}\textbf{93.1}\textsubscript{\textcolor{red}{$\uparrow$20.0}} & \cellcolor{lime}\textbf{94.1}\textsubscript{\textcolor{red}{$\uparrow$18.9}} & \cellcolor{lime}\textbf{92.6}\textsubscript{\textcolor{red}{$\uparrow$16.6}} \\
\hline
\end{tabular}
\end{adjustbox}
\caption{Performance comparison of SRLCG and other methods based on project completeness metrics, measured relative to the Vanilla LLM. Improvements are indicated in \textcolor{red}{red}, while declines are shown in \textcolor{blue}{blue} subscripts. The best score is highlighted in \textbf{bold}, and the second-best is \underline{underlined}.}
\label{tab:project_completeness}
\end{table*}

\noindent $\bullet$ \textbf{Implementation Settings}. We access GPT via the OpenAI API (\texttt{gpt-4-turbo-2024-04-09}) and \texttt{DeepSeek-Chat} (now DeepSeek-V3) through the DeepSeek API. For large-scale generation, we set the temperature to $0.3$. To ensure reliability, each data sample undergoes three experimental runs, with the average score used for evaluation. The base attenuation coefficient $\alpha$ is set to $0.1$, while the self-rectification frequency $f$ initializes at $0$ and increases with occurrences. The significance adjustment parameter $\beta$ is set to $1.2$. The details of feedback impact score $I^d$, $\mathcal{W}^d_{min}$ for each dimension $d$ can be found in Appendix~\ref{app:parameter_settings}. Further details, including MdCoT-DB module prompts, are provided in Appendix~\ref{sec:MdCoT-DB Prompts}, and Self-Rectification module prompts are provided in Appendix~\ref{sub:Self-Rectification Prompts}.

\subsection{Performance Comparison with Baselines}
\subsubsection{Code Length Evaluation}
\label{sec:code length}
As shown in Table~\ref{tab:model_comparison}, we measure the average byte size of all code files in the project directory, where longer code indicates better alignment with project requirements and feasibility. SRLCG significantly enhances code length across multiple domains. On DeepSeek-V3, it achieves an average length of $55,257.8$, surpassing Vanilla LLM by over \textbf{15$\times$}, while on GPT-4, it reaches $61,466.6$, exceeding Vanilla LLM by more than \textbf{16$\times$}. SRLCG surpasses other leading code generation methods, achieving an order-of-magnitude improvement.
These results underscore the effectiveness of SRLCG’s multidimensional CoT approach, which structures code generation across strategic, tactical, and operational dimensions, while also highlighting the efficacy of its dynamic backtracking algorithm in generating complete project code.

\subsubsection{Project Completeness Evaluation} \label{sec:project completeness} 

Table \ref{tab:project_completeness} presents the evaluation results for four project completeness metrics across different methods. For both DeepSeek-V3 and GPT-4, SRLCG outperforms all baselines and ranks highest in all metics. Specifically, SRLCG improves completeness by $16.5\%$, correctness by $13.1\%$, usability by $20.0\%$, and robustness by $19.1\%$ compared to Vanilla LLM in average of the DeepSeek-V3 and GPT-4. Overall, SRLCG achieves an average improvement of $16.9\%$, further validating its effectiveness in code generation tasks and demonstrating the success of our method's self-rectification mechanism. 

\setlength{\tabcolsep}{25pt} 
\begin{table}[t]
\centering
\begin{adjustbox}{width=1\linewidth, center} 
\fontsize{85}{95}\selectfont 
\begin{tabular}{c|c|c|c|c|c}
\toprule
\textbf{Backbone} & \textbf{Method} & \textbf{PS} & \textbf{TE(h)} & \textbf{CQI} & \textbf{TCR} \\
\midrule
\multirow{10}{*}{\centering\rotatebox{90}{DeepSeek-v3}} 
 & Vanilla LLM & 70.5 & 18.3  & 70.3 & 70.1 \\
 & CoT-pmt & 68.2\textsubscript{\textcolor{blue}{↓$$2.3$$}} & 17.7\textsubscript{\textcolor{blue}{↓$$0.6$$}} & 69.8\textsubscript{\textcolor{blue}{↓$$0.5$$}} & 69.5\textsubscript{\textcolor{blue}{↓$$0.6$$}} \\
 & KQG-CoT+ & 71.0\textsubscript{\textcolor{red}{↑$$0.5$$}} & 19.5\textsubscript{\textcolor{red}{↑$$1.2$$}} & 70.8\textsubscript{\textcolor{red}{↑$$0.5$$}} & 70.6\textsubscript{\textcolor{red}{↑$$0.5$$}} \\
 & SCOT(LI) & 69.8\textsubscript{\textcolor{blue}{↓$$0.7$$}} & 15.2\textsubscript{\textcolor{blue}{↓$$3.1$$}} & 69.5\textsubscript{\textcolor{blue}{↓$$0.8$$}} & 69.3\textsubscript{\textcolor{blue}{↓$$0.8$$}} \\
 & COTTON & 71.2\textsubscript{\textcolor{red}{↑$$0.7$$}} & 14.6\textsubscript{\textcolor{blue}{↓$$3.7$$}} & 70.9\textsubscript{\textcolor{red}{↑$$0.6$$}} & 70.7\textsubscript{\textcolor{red}{↑$$0.6$$}} \\
 & Self-planning & 69.0\textsubscript{\textcolor{blue}{↓$$1.5$$}} & 13.8\textsubscript{\textcolor{blue}{↓$$4.5$$}} & 68.8\textsubscript{\textcolor{blue}{↓$$1.5$$}} & 68.6\textsubscript{\textcolor{blue}{↓$$1.5$$}} \\
 & AceCoder & 70.8\textsubscript{\textcolor{red}{↑$$0.3$$}} & 20.0\textsubscript{\textcolor{red}{↑$$1.7$$}} & 70.5\textsubscript{\textcolor{red}{↑$$0.2$$}} & 70.3\textsubscript{\textcolor{red}{↑$$0.2$$}} \\
 & Scot(Md) & 69.5\textsubscript{\textcolor{blue}{↓$$1.0$$}} & 11.9\textsubscript{\textcolor{blue}{↓$$6.4$$}} & 69.2\textsubscript{\textcolor{blue}{↓$$1.1$$}} & 69.0\textsubscript{\textcolor{blue}{↓$$1.1$$}} \\
 & COP & 71.5\textsubscript{\textcolor{red}{↑$$1.0$$}} & 10.7\textsubscript{\textcolor{blue}{↓$$7.6$$}} & 71.2\textsubscript{\textcolor{red}{↑$$0.9$$}} & 71.0\textsubscript{\textcolor{red}{↑$$0.9$$}} \\
 & \cellcolor{lime}\textbf{SRLCG (ours)} & \cellcolor{lime}\textbf{94.6}\textsubscript{\textcolor{red}{↑$$24.1$$}} & \cellcolor{lime}\textbf{0.5}\textsubscript{\textcolor{blue}{↓$$17.8$$}} & \cellcolor{lime}\textbf{95.2}\textsubscript{\textcolor{red}{↑$$24.9$$}} & \cellcolor{lime}\textbf{95.4}\textsubscript{\textcolor{red}{↑$$25.3$$}} \\
\midrule
\multirow{10}{*}{\centering\rotatebox{90}{GPT-4}} 
 & Vanilla LLM & 70.2 & 15.6  & 70.5 & 70.3 \\
 & CoT-pmt & 69.0\textsubscript{\textcolor{blue}{↓$$1.2$$}} & 14.3\textsubscript{\textcolor{blue}{↓$$1.3$$}} & 70.0\textsubscript{\textcolor{blue}{↓$$0.5$$}} & 69.8\textsubscript{\textcolor{blue}{↓$$0.5$$}} \\
 & KQG-CoT+ & 70.8\textsubscript{\textcolor{red}{↑$$0.6$$}} & 17.0\textsubscript{\textcolor{red}{↑$$1.4$$}} & 70.7\textsubscript{\textcolor{red}{↑$$0.2$$}} & 70.5\textsubscript{\textcolor{red}{↑$$0.2$$}} \\
 & SCOT(Md) & 69.5\textsubscript{\textcolor{blue}{↓$$0.7$$}} & 12.4\textsubscript{\textcolor{blue}{↓$$3.2$$}} & 69.8\textsubscript{\textcolor{blue}{↓$$0.7$$}} & 69.6\textsubscript{\textcolor{blue}{↓$$0.7$$}} \\
 & COTTON & 71.0\textsubscript{\textcolor{red}{↑$$0.8$$}} & 18.5\textsubscript{\textcolor{red}{↑$$2.9$$}} & 70.9\textsubscript{\textcolor{red}{↑$$0.4$$}} & 70.7\textsubscript{\textcolor{red}{↑$$0.4$$}} \\
 & Self-planning & 68.8\textsubscript{\textcolor{blue}{↓$$1.4$$}} & 10.5\textsubscript{\textcolor{blue}{↓$$5.1$$}} & 69.0\textsubscript{\textcolor{blue}{↓$$1.5$$}} & 68.8\textsubscript{\textcolor{blue}{↓$$1.5$$}} \\
 & AceCoder & 70.5\textsubscript{\textcolor{red}{↑$$0.3$$}} & 19.0\textsubscript{\textcolor{red}{↑$$3.4$$}} & 70.4\textsubscript{\textcolor{blue}{↓$$0.1$$}} & 70.2\textsubscript{\textcolor{blue}{↓$$0.1$$}} \\
 & Scot(Md) & 69.2\textsubscript{\textcolor{blue}{↓$$1.0$$}} & 8.9\textsubscript{\textcolor{blue}{↓$$6.7$$}} & 69.5\textsubscript{\textcolor{blue}{↓$$1.0$$}} & 69.3\textsubscript{\textcolor{blue}{↓$$1.0$$}} \\
 & COP & 71.3\textsubscript{\textcolor{red}{↑$$1.1$$}} & 7.4\textsubscript{\textcolor{blue}{↓$$8.2$$}} & 71.0\textsubscript{\textcolor{red}{↑$$0.5$$}} & 70.8\textsubscript{\textcolor{red}{↑$$0.5$$}} \\
 & \cellcolor{lime}\textbf{SRLCG (ours)} & \cellcolor{lime}\textbf{96.8}\textsubscript{\textcolor{red}{↑$$26.6$$}} & \cellcolor{lime}\textbf{0.3}\textsubscript{\textcolor{blue}{↓$$15.3$$}} & \cellcolor{lime}\textbf{96.3}\textsubscript{\textcolor{red}{↑$$25.8$$}} & \cellcolor{lime}\textbf{97.5}\textsubscript{\textcolor{red}{↑$$27.2$$}} \\
\bottomrule
\end{tabular}
\end{adjustbox}
\caption{Performance comparison of SRLCG and other methods on human evaluation metrics, measured relative to the Vanilla LLM. Improvements are indicated in \textcolor{red}{red}, while declines are shown in \textcolor{blue}{blue} subscripts. The best score is highlighted in \textbf{bold}.}
\label{tab:model_comparison_extended}
\end{table}

\subsubsection{Human Evaluation}
\label{sec:human evaluation}

\begin{figure*}[t]
    \centering
    \begin{subfigure}{0.24\textwidth} 
        \centering
        \includegraphics[width=\linewidth]{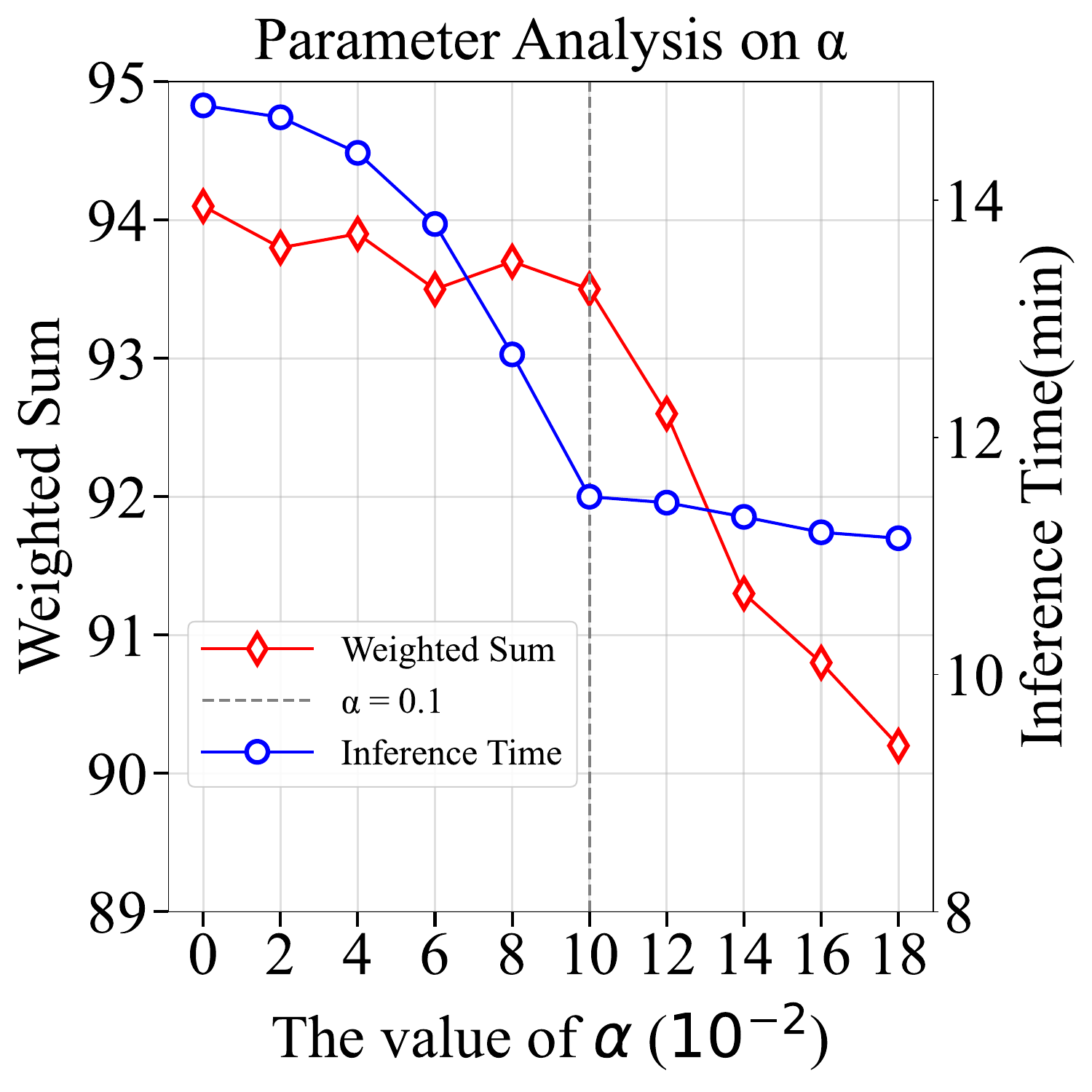}
        \caption{Analysis of $\alpha$}
        \label{fig:al1}
    \end{subfigure}
    \begin{subfigure}{0.24\textwidth}
        \centering
        \includegraphics[width=\linewidth]{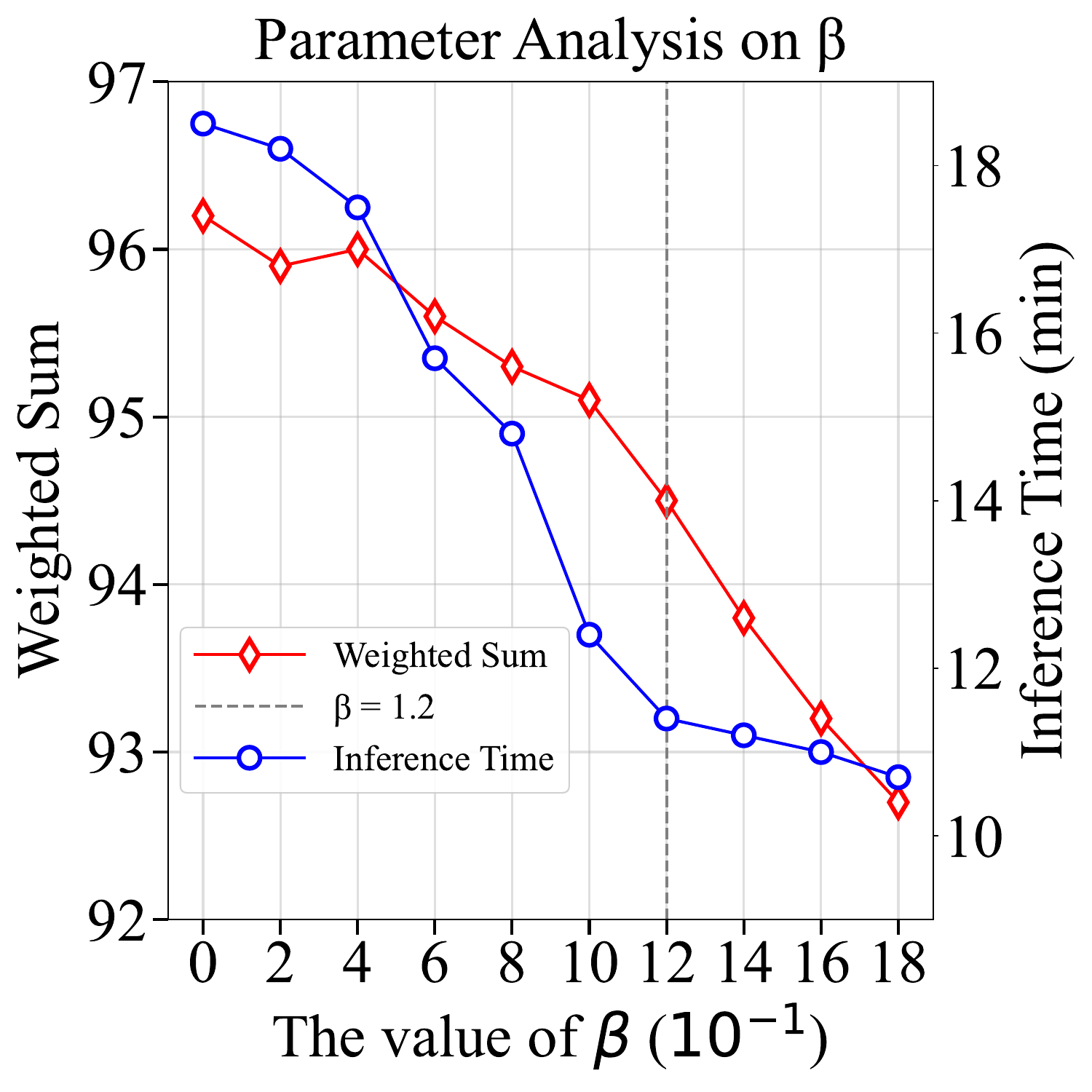}
        \caption{Analysis of $\beta$}
        \label{fig:al2}
    \end{subfigure}
    \begin{subfigure}{0.24\textwidth}
        \centering
        \includegraphics[width=\linewidth]{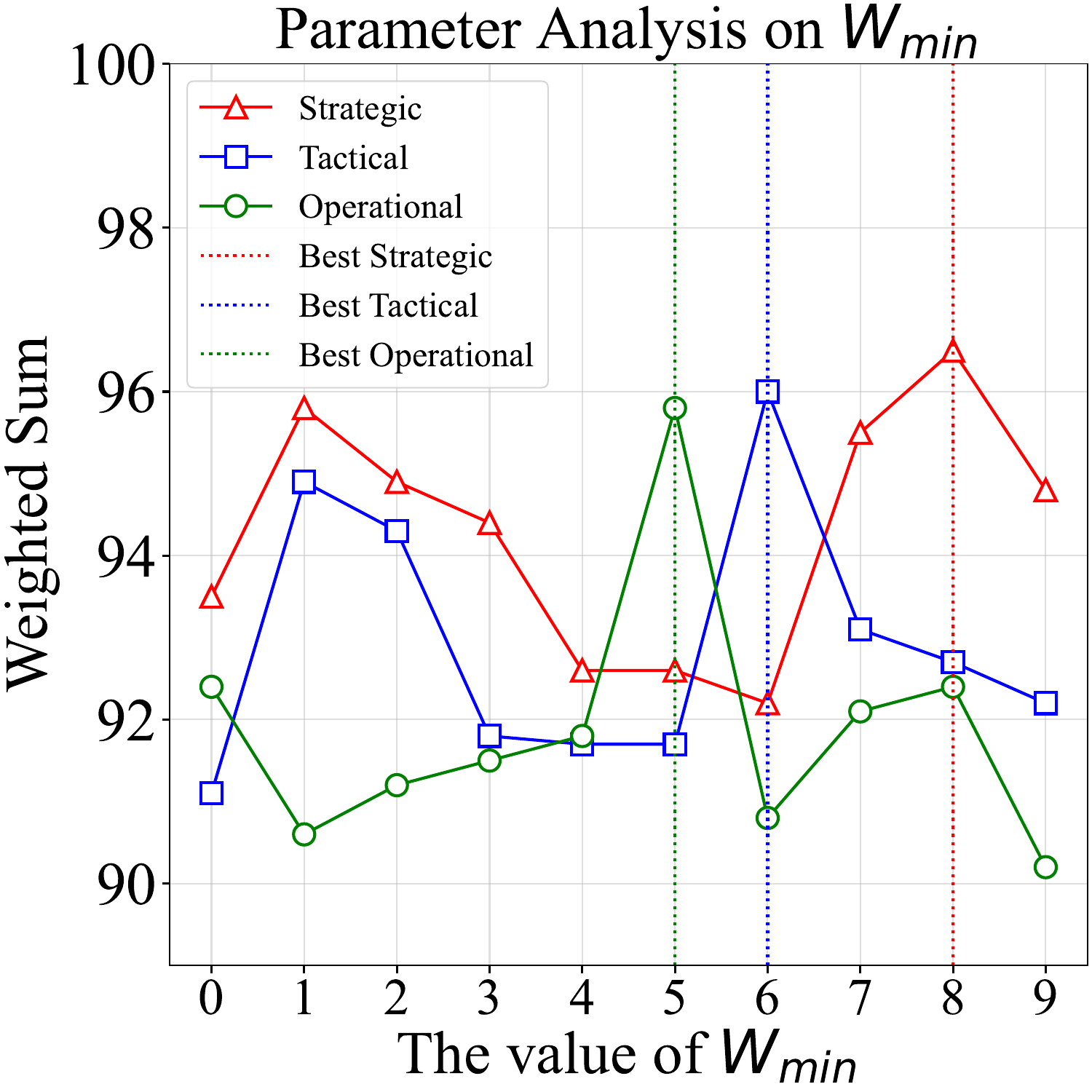}
        \caption{Analysis of $W_{min}$}
        \label{fig:al3}
    \end{subfigure}
    \begin{subfigure}{0.24\textwidth}
        \centering
        \includegraphics[width=\linewidth]{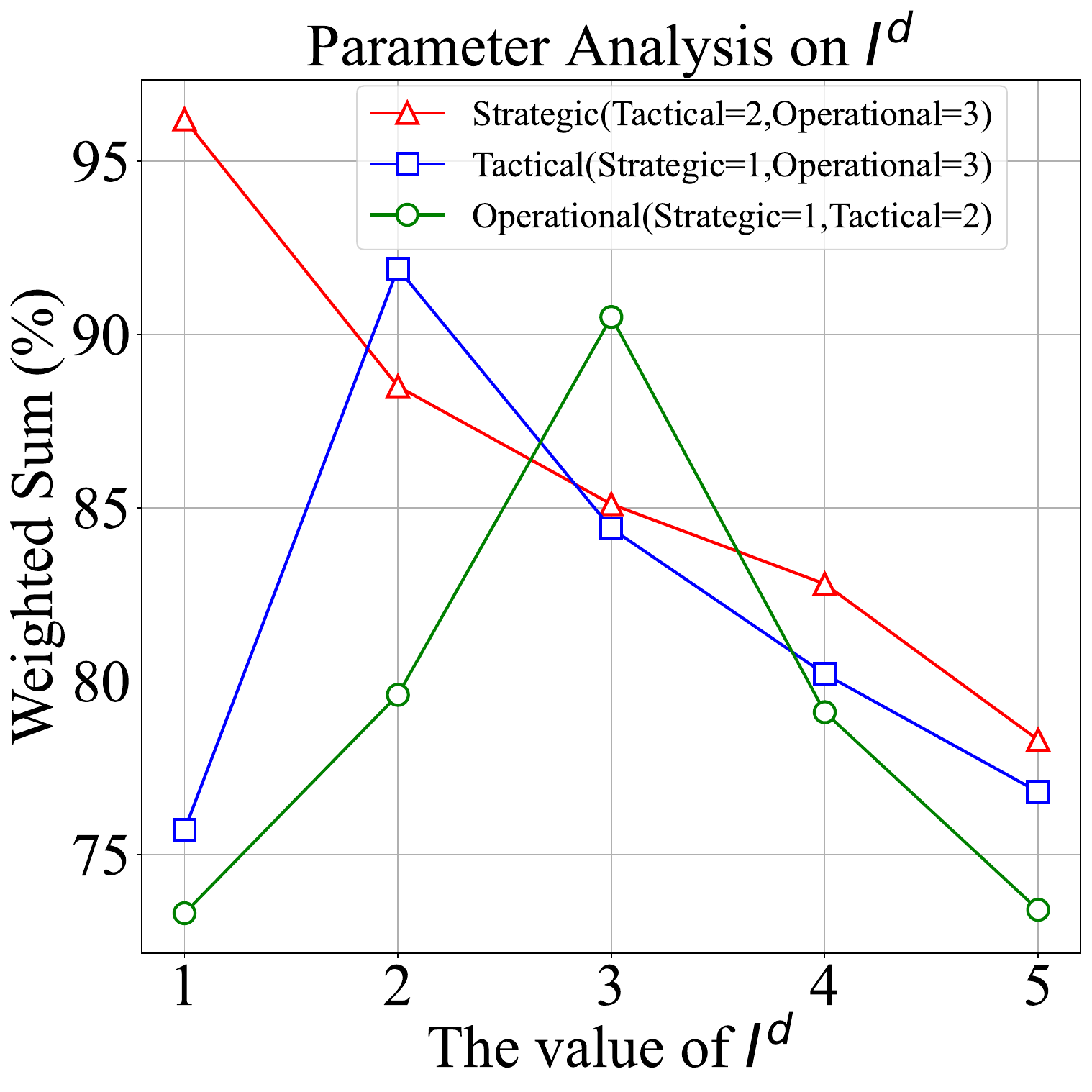}
        \caption{Analysis of $I^d$}
        \label{fig:al4}
    \end{subfigure}
    \caption{Analysis of parameters in the SRLCG framework.}
    \label{fig:tr}
\end{figure*}

As shown in Table~\ref{tab:model_comparison_extended}, We complement our evaluation with human evaluation by recruiting programmers with at least five years of experience. They evaluate the generated code based on four criteria: (1) \textit{Project Satisfaction} (PS), measuring alignment with project expectations; (2) \textit{Time Efficiency} (TE), the hours required to complete a functional project; (3) \textit{Code Quality Improvement} (CQI), assessing enhancements over the baseline; and (4) \textit{Task Completion Rate} (TCR), the percentage of successfully completed tasks.
Table~\ref{tab:model_comparison_extended} shows SRLCG improves PS by $25.4\%$, CQI by $25.3\%$, and Usability by $26.3\%$ over Vanilla LLM (avg. DeepSeek-V3 and GPT-4), while reducing develop time by $36\times$. 
These results highlight SRLCG's superiority in high-quality, efficient, and reliable large-scale code generation.

\subsection{Parameter Analysis}

\textbf{Parameter Analysis on $\alpha$ and $\beta$}: As shown in Figure~\hyperref[fig:tr]{\ref{fig:tr}(a)} and~\hyperref[fig:tr]{\ref{fig:tr}(b)}, as the value of the base attenuation coefficient $\alpha$ increases, both the inference time and the Weighted Sum value continuously decrease. This is because once the value of $\alpha$ rises, the probability of each module undergoing verification decreases, leading to a reduction in both metrics. We can observe that after $\alpha=0.1$, the inference time decreases slowly while the Weighted Sum value drops sharply. Therefore, $\alpha=0.1$ is the optimal setting. A similar trend is observed for $\beta$, with the best setting being $1.2$.

\textbf{Parameter Analysis on $\mathcal{W}^d_\text{min}$}: As shown in Figure~\hyperref[fig:tr]{\ref{fig:tr}(c)}, for $\mathcal{W}^d_\text{min}$, we test values ranging from $0$ to $1$ across the three dimensions. The results show that the Weighted Sum reaches its highest value when $\mathcal{W}^d_\text{min}$ is set to $0.8$ for strategic dimension, $0.6$ for tactical dimension, and $0.5$ for operational dimension.

\textbf{Parameter Analysis on $I^d$}: As shown in Figure~\hyperref[fig:tr]{\ref{fig:tr}(d)}, $I^d$ represents the feedback impact score of dimension $d$. When $I^d$ exceeds $3$, we observe that the self-rectification process is almost never executed. This is because an excessively large $I^d$ results in a significantly small $\mathcal{W}_{\text{current}}$, causing the self-rectification to be skipped entirely. Therefore, when analyzing one dimension, we keep the other two dimensions fixed, such as in the case of strategic dimension $I^{\text{strategic}}$ analysis where we fix $I^{\text{tactical}}=2$ and $I^{\text{operational}}=3$.
Through this analysis, we find that the optimal settings are $I^{\text{strategic}}=1$, $I^{\text{tactical}}=2$, and $I^{\text{operational}}=3$.

\subsection{Ablation Study}
To analyze the rationality and the effectiveness of
the designed modules in our SRLCG framework, we conduct an ablation study to compare SRLCG with its variants in Appendix~\ref{sec:ablation}.

\subsection{Case Study}
To clearly demonstrate the practicality of SRLCG, we conduct a case study on project management tool development. The detailed project development prompt and a subset of the code generated by SRLCG are provided in Appendix~\ref{sec:case study}.

\section{Conclusion}
In this paper, we propose SRLCG to enable users with limited programming knowledge to generate complex large-scale code. Unlike prior work on short code generation, we introduce a Multidimensional CoT with dynamic backtracking, tailored for long-code generation. Our self-rectification module refines rationale at each dimension, ensuring the code is comprehensive, accurate, and robust.
Extensive experiments demonstrate the effectiveness and practicality of SRLCG compared to leading existing code generation methods for large-scale code generation. We hope our work can be applied across various domains where users require the generation of complex large-scale project code.
\section*{Limitations}
Since the MdCoT-DB module in our SRLCG framework is built upon multidimensional CoT, which inherently depends on LLM inference, the associated processing time poses a persistent challenge. Given the nature of our task in large-scale code generation, the framework's average processing time for a single project is approximately eleven minutes on two Tesla V100 GPUs (32GB each) and tends to increase as project complexity grows. This processing time may exceed user expectations. Future work will explore methods to enhance inference efficiency and reduce overall processing overhead.
\bibliography{acl_latex}

\begin{thebibliography}{33}
\providecommand{\natexlab}[1]{#1}

\bibitem[{Adnan et~al.(2025)Adnan, Xu, and Kuhn}]{adnan2025largelanguagemodelguided}
Muntasir Adnan, Zhiwei Xu, and Carlos C.~N. Kuhn. 2025.
\newblock \href {https://arxiv.org/abs/2502.02928} {Large language model guided self-debugging code generation}.
\newblock \emph{Preprint}, arXiv:2502.02928.

\bibitem[{Chen et~al.(2024{\natexlab{a}})Chen, Tang, Chu, Chen, Wang, Liu, and Qin}]{chen2024divideandconquer}
Jingchang Chen, Hongxuan Tang, Zheng Chu, Qianglong Chen, Zekun Wang, Ming Liu, and Bing Qin. 2024{\natexlab{a}}.
\newblock Divide-and-conquer meets consensus: Unleashing the power of functions in code generation.
\newblock In \emph{Proceedings of the 38th Annual Conference on Neural Information Processing Systems}.

\bibitem[{Chen et~al.(2024{\natexlab{b}})Chen, Lin, Sch{\"a}rli, and Zhou}]{chen2024teaching}
Xinyun Chen, Maxwell Lin, Nathanael Sch{\"a}rli, and Denny Zhou. 2024{\natexlab{b}}.
\newblock Teaching large language models to self-debug.
\newblock In \emph{Proceedings of the 20th International Conference on Learning Representations}.

\bibitem[{DeepSeek-AI(2024)}]{deepseekai2024deepseekv3technicalreport}
DeepSeek-AI. 2024.
\newblock \href {https://arxiv.org/abs/2412.19437} {Deepseek-v3 technical report}.
\newblock \emph{Preprint}, arXiv:2412.19437.

\bibitem[{Du et~al.(2024)Du, Liu, Wang, Wang, Liu, Chen, Feng, Sha, Peng, and Lou}]{Du2024Class}
Xueying Du, Mingwei Liu, Kaixin Wang, Hanlin Wang, Junwei Liu, Yixuan Chen, Jiayi Feng, Chaofeng Sha, Xin Peng, and Yiling Lou. 2024.
\newblock Evaluating large language models in class-level code generation.
\newblock In \emph{Proceedings of the 46th IEEE/ACM International Conference on Software Engineering}, page 1–13.

\bibitem[{Huang et~al.(2024)Huang, Bu, Qing, and Cui}]{huang2024codecottacklingcodesyntax}
Dong Huang, Qingwen Bu, Yuhao Qing, and Heming Cui. 2024.
\newblock \href {https://arxiv.org/abs/2308.08784} {Codecot: Tackling code syntax errors in cot reasoning for code generation}.
\newblock \emph{Preprint}, arXiv:2308.08784.

\bibitem[{Huang et~al.(2025)Huang, Song, Wang, Zhao, Chen, Juefei-Xu, and Ma}]{Huang_2025}
Yuheng Huang, Jiayang Song, Zhijie Wang, Shengming Zhao, Huaming Chen, Felix Juefei-Xu, and Lei Ma. 2025.
\newblock Look before you leap: An exploratory study of uncertainty analysis for large language models.
\newblock \emph{IEEE Transactions on Software Engineering}, page 1–18.

\bibitem[{Jiang et~al.(2024{\natexlab{a}})Jiang, Li, Wang, Zhou, Hossain, Ray, Kumar, Ma, and Deoras}]{jiang2024ledex}
Nan Jiang, Xiaopeng Li, Shiqi Wang, Qiang Zhou, Soneya~Binta Hossain, Baishakhi Ray, Varun Kumar, Xiaofei Ma, and Anoop Deoras. 2024{\natexlab{a}}.
\newblock Ledex: Training {LLM}s to better self-debug and explain code.
\newblock In \emph{Proceedings of the 38th Annual Conference on Neural Information Processing Systems}.

\bibitem[{Jiang et~al.(2024{\natexlab{b}})Jiang, Dong, Wang, Fang, Shang, Li, Jin, and Jiao}]{jiang2024self}
Xue Jiang, Yihong Dong, Lecheng Wang, Zheng Fang, Qiwei Shang, Ge~Li, Zhi Jin, and Wenpin Jiao. 2024{\natexlab{b}}.
\newblock Self-planning code generation with large language models.
\newblock \emph{ACM Transactions on Software Engineering and Methodology}, (7):1--30.

\bibitem[{Le et~al.(2024)Le, Chen, Saha, Gokul, Sahoo, and Joty}]{le2023codechain}
Hung Le, Hailin Chen, Amrita Saha, Akash Gokul, Doyen Sahoo, and Shafiq Joty. 2024.
\newblock Codechain: Towards modular code generation through chain of self-revisions with representative sub-modules.
\newblock In \emph{Proceeding of the 20th International Conference on Learning Representations}.

\bibitem[{Li et~al.(2025)Li, Li, Li, and Jin}]{Li2025Structured}
Jia Li, Ge~Li, Yongmin Li, and Zhi Jin. 2025.
\newblock Structured chain-of-thought prompting for code generation.
\newblock \emph{ACM Transactions on Software Engineering and Methodology}, (2):1--23.

\bibitem[{Li et~al.(2024)Li, Zhao, Li, Li, and Jin}]{Li2024AceCoder}
Jia Li, Yunfei Zhao, Yongmin Li, Ge~Li, and Zhi Jin. 2024.
\newblock Acecoder: An effective prompting technique specialized in code generation.
\newblock \emph{ACM Transactions on Software Engineering and Methodology}, 33(8):1--26.

\bibitem[{Liang et~al.(2023)Liang, Wang, Zhu, Wang, Qian, and Lan}]{liang2023prompting}
Yuanyuan Liang, Jianing Wang, Hanlun Zhu, Lei Wang, Weining Qian, and Yunshi Lan. 2023.
\newblock Prompting large language models with chain-of-thought for few-shot knowledge base question generation.
\newblock In \emph{Proceedings of the 2023 Conference on Empirical Methods in Natural Language Processing}, pages 4329--4343.

\bibitem[{Ling et~al.(2023)Ling, Fang, Li, Huang, Lee, Memisevic, and Su}]{ling2023deductive}
Zhan Ling, Yunhao Fang, Xuanlin Li, Zhiao Huang, Mingu Lee, Roland Memisevic, and Hao Su. 2023.
\newblock Deductive verification of chain-of-thought reasoning.
\newblock In \emph{Procddings of the 37th Conference on Neural Information Processing Systems}.

\bibitem[{OpenAI(2024)}]{openai2024gpt4technicalreport}
OpenAI. 2024.
\newblock \href {https://arxiv.org/abs/2303.08774} {{GPT}-4 technical report}.
\newblock \emph{Preprint}, arXiv:2303.08774.

\bibitem[{Saxena et~al.(2024)Saxena, Chopra, and Tripathi}]{saxena2024evaluatingconsistencyreasoningcapabilities}
Yash Saxena, Sarthak Chopra, and Arunendra~Mani Tripathi. 2024.
\newblock \href {https://arxiv.org/abs/2404.16478} {Evaluating consistency and reasoning capabilities of large language models}.
\newblock \emph{Preprint}, arXiv:2404.16478.

\bibitem[{Stechly et~al.(2024)Stechly, Valmeekam, and Kambhampati}]{stechly2024chainthoughtlessnessanalysiscot}
Kaya Stechly, Karthik Valmeekam, and Subbarao Kambhampati. 2024.
\newblock \href {https://arxiv.org/abs/2405.04776} {Chain of thoughtlessness? an analysis of cot in planning}.
\newblock \emph{Preprint}, arXiv:2405.04776.

\bibitem[{Sultan et~al.(2024)Sultan, Ganhotra, and Astudillo}]{Md2024Scot}
Md~Arafat Sultan, Jatin Ganhotra, and Ramón~Fernandez Astudillo. 2024.
\newblock \href {https://arxiv.org/abs/2402.11770} {Structured chain-of-thought prompting for few-shot generation of content-grounded {QA} conversations}.
\newblock \emph{Preprint}, arXiv:2402.11770.

\bibitem[{Tian et~al.(2025)Tian, Yan, Yang, Zhao, Chen, Wang, Luo, Ma, and Song}]{tian2025codehaluinvestigatingcodehallucinations}
Yuchen Tian, Weixiang Yan, Qian Yang, Xuandong Zhao, Qian Chen, Wen Wang, Ziyang Luo, Lei Ma, and Dawn Song. 2025.
\newblock \href {https://arxiv.org/abs/2405.00253} {Codehalu: Investigating code hallucinations in llms via execution-based verification}.
\newblock \emph{Preprint}, arXiv:2405.00253.

\bibitem[{Wang et~al.(2024{\natexlab{a}})Wang, Zhou, Song, Huang, Chen, Ma, and Zhang}]{wang2024largelanguagemodelsfail}
Zhijie Wang, Zijie Zhou, Da~Song, Yuheng Huang, Shengmai Chen, Lei Ma, and Tianyi Zhang. 2024{\natexlab{a}}.
\newblock \href {https://arxiv.org/abs/2406.08731} {Where do large language models fail when generating code?}
\newblock \emph{Preprint}, arXiv:2406.08731.

\bibitem[{Wang et~al.(2024{\natexlab{b}})Wang, Chu, Doan, Ni, Yang, and Zhang}]{wang2024historydevelopmentprincipleslarge}
Zichong Wang, Zhibo Chu, Thang~Viet Doan, Shiwen Ni, Min Yang, and Wenbin Zhang. 2024{\natexlab{b}}.
\newblock \href {https://arxiv.org/abs/2402.06853} {History, development, and principles of large language models-an introductory survey}.
\newblock \emph{Preprint}, arXiv:2402.06853.

\bibitem[{Wei et~al.(2022)Wei, Wang, Schuurmans, Bosma, ichter, Xia, Chi, Le, and Zhou}]{Wei2022Chain-of-thoughtprompting}
Jason Wei, Xuezhi Wang, Dale Schuurmans, Maarten Bosma, brian ichter, Fei Xia, Ed~Chi, Quoc~V Le, and Denny Zhou. 2022.
\newblock Chain-of-thought prompting elicits reasoning in large language models.
\newblock In \emph{Proceedings of the Advances in Neural Information Processing Systems}, pages 24824--24837.

\bibitem[{Weir et~al.(2024)Weir, Khalifa, Qiu, Weller, and Clark}]{Weir2024Learning}
Nathaniel Weir, Muhammad Khalifa, Linlu Qiu, Orion Weller, and Peter Clark. 2024.
\newblock Learning to reason via program generation, emulation, and search.
\newblock In \emph{Proceedings of the Advances in Neural Information Processing Systems}, pages 36390--36413.

\bibitem[{Wen et~al.(2025)Wen, Zhu, Liu, Ren, Du, and Yan}]{wen2025fixingfunctionlevelcodegeneration}
Hao Wen, Yueheng Zhu, Chao Liu, Xiaoxue Ren, Weiwei Du, and Meng Yan. 2025.
\newblock \href {https://arxiv.org/abs/2409.00676} {Fixing function-level code generation errors for foundation large language models}.
\newblock \emph{Preprint}, arXiv:2409.00676.

\bibitem[{Xie et~al.(2024)Xie, Guo, Yu, and Li}]{xie2024Calibrating}
Zhihui Xie, Jizhou Guo, Tong Yu, and Shuai Li. 2024.
\newblock Calibrating reasoning in language models with internal consistency.
\newblock In \emph{Proceedings of the Advances in Neural Information Processing Systems}, pages 114872--114901.

\bibitem[{Xu et~al.(2024)Xu, Li, Wu, Wei, Du, Wang, and Song}]{Xu2024COP}
Bo~Xu, Shufei Li, Yifei Wu, Shouang Wei, Ming Du, Hongya Wang, and Hui Song. 2024.
\newblock Chain-of-program prompting with open-source large language models for text-to-sql.
\newblock In \emph{Proceedings of the 2024 International Joint Conference on Neural Networks (IJCNN)}, pages 1--8.

\bibitem[{Yang et~al.(2024)Yang, Zhou, Chen, Zhang, Zhuo, and Chen}]{Yang2024COTTON}
Guang Yang, Yu~Zhou, Xiang Chen, Xiangyu Zhang, Terry~Yue Zhuo, and Taolue Chen. 2024.
\newblock Chain-of-thought in neural code generation: From and for lightweight language models.
\newblock \emph{IEEE Transactions on Software Engineering}, (9):2437--2457.

\bibitem[{Yen et~al.(2024)Yen, Zhu, Suh, Xia, and Zhao}]{Yen2024CoLadder}
Ryan Yen, Jiawen~Stefanie Zhu, Sangho Suh, Haijun Xia, and Jian Zhao. 2024.
\newblock Coladder: Manipulating code generation via multi-level blocks.
\newblock In \emph{Proceedings of the 37th Annual ACM Symposium on User Interface Software and Technology}, pages 1--20.

\bibitem[{Zhang et~al.(2023)Zhang, Li, Li, Li, and Jin}]{zhang2023selfeditfaultawarecodeeditor}
Kechi Zhang, Zhuo Li, Jia Li, Ge~Li, and Zhi Jin. 2023.
\newblock \href {https://arxiv.org/abs/2305.04087} {Self-edit: Fault-aware code editor for code generation}.
\newblock \emph{Preprint}, arXiv:2305.04087.

\bibitem[{Zhao et~al.(2024{\natexlab{a}})Zhao, Zhou, Li, Tang, Wang, Hou, Min, Zhang, Zhang, Dong, Du, Yang, Chen, Chen, Jiang, Ren, Li, Tang, Liu, Liu, Nie, and Wen}]{zhao2024surveylargelanguagemodels}
Wayne~Xin Zhao, Kun Zhou, Junyi Li, Tianyi Tang, Xiaolei Wang, Yupeng Hou, Yingqian Min, Beichen Zhang, Junjie Zhang, Zican Dong, Yifan Du, Chen Yang, Yushuo Chen, Zhipeng Chen, Jinhao Jiang, Ruiyang Ren, Yifan Li, Xinyu Tang, Zikang Liu, Peiyu Liu, Jian-Yun Nie, and Ji-Rong Wen. 2024{\natexlab{a}}.
\newblock \href {https://arxiv.org/abs/2303.18223} {A survey of large language models}.
\newblock \emph{Preprint}, arXiv:2303.18223.

\bibitem[{Zhao et~al.(2024{\natexlab{b}})Zhao, Monti, Lehmann, and Assem}]{zhao2024enhancing}
Zheng Zhao, Emilio Monti, Jens Lehmann, and Haytham Assem. 2024{\natexlab{b}}.
\newblock Enhancing contextual understanding in large language models through contrastive decoding.
\newblock In \emph{Proceedings of the 2024 Conference of the North American Chapter of the Association for Computational Linguistics: Human Language Technologies}, pages 4225--4237.

\bibitem[{Zheng et~al.(2023)Zheng, Sharan, Jaiswal, Wang, Xi, Xu, and Wang}]{Zheng2023Outline}
Wenqing Zheng, S~P Sharan, Ajay~Kumar Jaiswal, Kevin Wang, Yihan Xi, Dejia Xu, and Zhangyang Wang. 2023.
\newblock Outline, then details: syntactically guided coarse-to-fine code generation.
\newblock In \emph{Proceedings of the 40th International Conference on Machine Learning}.

\bibitem[{Zhou et~al.(2024)Zhou, Hwang, Ren, and Sap}]{zhou2024relying}
Kaitlyn Zhou, Jena Hwang, Xiang Ren, and Maarten Sap. 2024.
\newblock Relying on the unreliable: The impact of language models' reluctance to express uncertainty.
\newblock In \emph{Proceedings of the 62nd Annual Meeting of the Association for Computational Linguistics}, pages 3623--3643.

\end{thebibliography}
\newpage
\appendix
\section{Dataset}
\label{sec:Dataset}
\subsection{Data Format in the Dataset}
\label{sec:appendix_dataset}
Our large-scale code generation dataset consists of individual prompts, each representing a real-world scenario where users need to generate project code based on practical requirements. These prompts are formatted according to the following structure.

\textbf{Task Definition:}
The task definition section provides a detailed description of the project's goals, functional requirements, and expected outcomes. It clarifies the type of application to be developed, its core functionalities, and the user's expectations, offering a clear context for code generation.

\textbf{Key Features:}
The key features section further elaborates on the specific functional requirements of the task, listing the core functional modules and their detailed descriptions. Each module includes specific implementation requirements, such as input-output formats, state management, and user interaction logic.

\textbf{Technical Specifications:}
The technical specifications section defines the programming languages, development environments, and other technical requirements needed to implement the task. This section provides technical constraints for code generation, ensuring that the generated code meets practical development needs.

\subsection{Applications of the Dataset}
This dataset can be used to train and evaluate code generation models, particularly for long code generation tasks. By providing detailed task descriptions and functional requirements, the dataset enables models to generate high-quality code that aligns with real-world application scenarios. Additionally, its coverage of multiple domains makes it suitable for a wide range of code generation research.
\section{Evaluation Metrics}
\label{sec:evaluation_metrics}
In this section, we provide detailed descriptions of the evaluation metrics used in our study.

\subsection{Code Length}
Code Length is used to calculate the average length of all generated code in a project. This metric evaluates the total byte size of code files within the project folder. The final metric value is the average of the Code Length values across all samples.

\subsection{Project Completeness}
Project Completeness measures the overall completeness of projects generated by the CoT method. Specifically, it includes four sub-metrics:
\begin{itemize}
    \item \textbf{Completeness}: The degree to which the project is fully implemented.
    \item \textbf{Correctness}: The accuracy of the generated code.
    \item \textbf{Usability}: The ease of use and practicality of the code.
    \item \textbf{Robustness}: The resilience of the code to errors and edge cases.
\end{itemize}
To evaluate these metrics, we designed unique prompts for each sub-metric. The evaluation process involves providing the generated project code and the corresponding prompts to a large language model, which then assigns scores. The final score for each metric is the average of scores across all project code files.

\subsection{Human Evaluation}
We recruited programmers with at least five years of experience to evaluate the code generated by different methods. The evaluation covers four dimensions:
\begin{itemize}
    \item \textbf{Project Satisfaction (PS)}: Measures alignment with project expectations.
    \item \textbf{Time Efficiency (TE)}: Evaluates the hours required to complete a functional project.
    \item \textbf{Code Quality Improvement (CQI)}: Assesses enhancements over the baseline.
    \item \textbf{Task Completion Rate (TCR)}: Measures the percentage of successfully completed tasks.
\end{itemize}
The scores for these dimensions are obtained from the programmers' evaluations.

\section{Baseline Details}

\label{sec:baselines}
\begin{table*}[ht]
\centering
\Huge
\resizebox{\linewidth}{!}{ 
\begin{tabular}{c|l|c|c|c|c|c}
\toprule
\textbf{Model} & \textbf{Methods} & \textbf{Completeness} & \textbf{Correctness} & \textbf{Usability} & \textbf{Robustness} & \textbf{Weighted Sum} \\
\midrule
\multirow{5}{*}{DeepSeek-V3} 
 & SRLCG & \textbf{96.7} & \textbf{94.6} & \textbf{95.2} & \textbf{95.4} & \textbf{95.5} \\
 & SRLCG w/o DB & 75.3\textsubscript{\(\textcolor{blue}{\mathbin{\downarrow} 21.4}\)} & 79.5\textsubscript{\(\textcolor{blue}{\mathbin{\downarrow} 25.1}\)} & 83.7\textsubscript{\(\textcolor{blue}{\mathbin{\downarrow} 11.5}\)} & 85.5\textsubscript{\(\textcolor{blue}{\mathbin{\downarrow} 9.9}\)} & 81.0\textsubscript{\(\textcolor{blue}{\mathbin{\downarrow} 14.5}\)} \\
 & SRLCG w/o SV & 86.6\textsubscript{\(\textcolor{blue}{\mathbin{\downarrow} 10.1}\)} & 92.2\textsubscript{\(\textcolor{blue}{\mathbin{\downarrow} 2.4}\)} & 90.1\textsubscript{\(\textcolor{blue}{\mathbin{\downarrow} 5.1}\)} & 91.3\textsubscript{\(\textcolor{blue}{\mathbin{\downarrow} 4.1}\)} & 90.1\textsubscript{\(\textcolor{blue}{\mathbin{\downarrow} 5.4}\)} \\
 & SRLCG w/o PA & \underline{92.2}\textsubscript{\(\textcolor{blue}{\mathbin{\downarrow} 4.5}\)} & \underline{93.5}\textsubscript{\(\textcolor{blue}{\mathbin{\downarrow} 1.1}\)} & \underline{93.0}\textsubscript{\(\textcolor{blue}{\mathbin{\downarrow} 2.2}\)} & \underline{93.8}\textsubscript{\(\textcolor{blue}{\mathbin{\downarrow} 1.6}\)} & \underline{93.1}\textsubscript{\(\textcolor{blue}{\mathbin{\downarrow} 2.4}\)} \\
\midrule
\multirow{5}{*}{GPT-4} 
 & SRLCG & \textbf{98.1} & \textbf{95.2} & \textbf{96.3} & \textbf{97.5} & \textbf{96.8} \\
 & SRLCG w/o DB & 78.6\textsubscript{\(\textcolor{blue}{\mathbin{\downarrow} 19.5}\)} & 83.4\textsubscript{\(\textcolor{blue}{\mathbin{\downarrow} 11.8}\)} & 87.5\textsubscript{\(\textcolor{blue}{\mathbin{\downarrow} 8.8}\)} & 88.0\textsubscript{\(\textcolor{blue}{\mathbin{\downarrow} 9.5}\)} & 84.4\textsubscript{\(\textcolor{blue}{\mathbin{\downarrow} 12.4}\)} \\
 & SRLCG w/o SV & 87.4\textsubscript{\(\textcolor{blue}{\mathbin{\downarrow} 10.7}\)} & 93.0\textsubscript{\(\textcolor{blue}{\mathbin{\downarrow} 2.2}\)} & 91.2\textsubscript{\(\textcolor{blue}{\mathbin{\downarrow} 5.1}\)} & 92.4\textsubscript{\(\textcolor{blue}{\mathbin{\downarrow} 5.1}\)} & 91.0\textsubscript{\(\textcolor{blue}{\mathbin{\downarrow} 5.8}\)} \\
 & SRLCG w/o PA & \underline{93.0}\textsubscript{\(\textcolor{blue}{\mathbin{\downarrow} 5.1}\)} & \underline{94.2}\textsubscript{\(\textcolor{blue}{\mathbin{\downarrow} 1.0}\)} & \underline{94.0}\textsubscript{\(\textcolor{blue}{\mathbin{\downarrow} 2.3}\)} & \underline{94.8}\textsubscript{\(\textcolor{blue}{\mathbin{\downarrow} 2.7}\)} & \underline{94.0}\textsubscript{\(\textcolor{blue}{\mathbin{\downarrow} 2.8}\)} \\
\bottomrule
\end{tabular}
}
\caption{Performance comparison of SRLCG with its variants. Declines are shown in \textcolor{blue}{blue} subscripts. The best score is highlighted in \textbf{bold}, and the second-best is \underline{underlined}.}
\label{tab:Ablation}
\end{table*}

We adopt state-of-the-art code generation methods as baselines, detailed as follows.
\label{sec:baseline_details}
\begin{itemize}
    \setlength{\itemindent}{0pt}  
    \setlength{\itemsep}{0pt}  
    \item \textbf{Vanilla LM},directly uses simple COT for reasoning, predicting the outcomes of the questions through in-context learning.
    \item \textbf{COTTON}, which focuses on chain-of-thought reasoning in neural code generation for lightweight language models.  
    \item \textbf{Self-planning}, a self-planning code generation method leveraging large language models without explicit context. 
    \item \textbf{AceCoder}, a specialized prompting technique for code generation.
    \item \textbf{Chain-of-thought prompting}, a foundational work in CoT reasoning. 
    \item \textbf{KQG-CoT+}, which applies CoT to few-shot knowledge base question generation.  
    \item \textbf{Scot(Md)}, which employs structured CoT for few-shot content-grounded QA generation. 
    \item \textbf{COP}, a chain-of-program prompting approach for text-to-SQL tasks. 
    \item \textbf{SCOT(LI)} , which applies structured CoT to code generation. 
\end{itemize}

\section{Parameter Settings}
\label{app:parameter_settings}

This section provides the parameter settings for the feedback impact score ($I^d$) and the minimum weight threshold ($\mathcal{W}^d_{\text{min}}$) across different dimensions.

\begin{table}[htbp]
\centering

\label{tab:parameter_settings}
\small
\begin{tabular}{|c|c|}
\hline
\textbf{Dimension} & \textbf{ $I^d$}  \\ \hline
Strategic Dimension & 1  \\ \hline
Tactical Dimension  & 2  \\ \hline
Operational Dimension & 3 \\ \hline
\end{tabular}
\caption{Parameter settings for feedback impact score ($I^d$).}
\end{table}

\begin{table}[htbp]
\label{tab:parameter_settings}
\small
\begin{tabular}{|c|c|}
\hline
\textbf{Dimension} & \textbf{ $\mathcal{W}^d_{\text{min}}$}  \\ \hline
Strategic Dimension & 0.8  \\ \hline
Tactical Dimension  & 0.6  \\ \hline
Operational Dimension & 0.5 \\ \hline
\end{tabular}
\centering
\caption{Parameter settings for minimum weight threshold ($\mathcal{W}^d_{\text{min}}$).}
\end{table}


\section{Ablation Study}
\label{sec:ablation}

In this section, we evaluate the impact of the dynamic backtracking algorithm in the MdCoT-DB module of SRLCG, the Self-Verification module, and Progressive Attenuation in the Self-Verification module on SRLCG's performance. The experiments, conducted using DeepSeek-V3 and GPT-4, assess Completeness, Correctness, Usability, Robustness, and Weighted Sum. We perform an ablation study on SRLCG and its three variants.
(1) \textit{SRLCG w/o DB}: SRLCG without the dynamic backtracking algorithm in the MdCoT-DB module.
(2) \textit{SRLCG W/o SR}: SRLCG without self-rectification module. (3) \textit{SRLCG w/o PA}: SRLCG without progressive attenuation in the self-rectification module.

Table~\ref{tab:Ablation} illustrates that the dynamic backtracking algorithm in the MdCoT module, the self-rectification module, and progressive attenuation in the self-rectification module are crucial for the model's performance. These results yield three key findings:
(1) \textbf{Removing dynamic backtracking algorithm in the MdCoT module causes a significant decline in weighted Sum} ($14.5\%$ on DeepSeek-V3, $12.4\%$ on GPT-4), highlighting its importance in ensuring comprehensive code generation.
(2) \textbf{Removing self-rectification module causes a decline in weighted Sum} ($5.4\%$ on DeepSeek-V3, $5.8\%$ on GPT-4), demonstrates the effectiveness of Self-Verification.
(3) \textbf{Omitting progressive attenuation in self-rectification module leads to moderate declines across all metrics}, confirming its role in stabilizing performance and maintaining consistency.
These findings underscore the critical role of all components in preserving robust and high-quality performance.

\clearpage

\section{MdCoT-DB Prompts}
\label{sec:MdCoT-DB Prompts}
\subsection{Strategic Dimension Prompt}
\begin{tcolorbox}[colframe=black, colback=white, arc=3mm, boxrule=1pt, width=\textwidth]
    \textbf{Strategic Dimension Prompt}  
    \tcblower
    \textbf{Task Definition:} 
    
    [User project requirement Prompt $\mathcal{P}$]  
    
    Please decompose the above task into several macro-level modules and briefly describe the responsibilities of each module. 
    
    \textbf{Format Requirements:}
    
    (1) Use JSON format.  
    
    (2) Each module should include the module name and a brief description of its responsibilities.
    
    \textbf{Example Output:}
    
    \begin{verbatim}
[
    {
        "Module": "[Module Name]",
        "Responsibility": "[Brief description of the module's function]"
    },
    {
        "Module": "[Module Name]",
        "Responsibility": "[Brief description of the module's function]"
    },
    {
        "Module": "[Module Name]",
        "Responsibility": "[Brief description of the module's function]"
    }
]
    \end{verbatim}

    \medskip
    
    \textbf{Outputs:} 
\end{tcolorbox}

\clearpage

\subsection{Tactical Dimension Prompt}
\begin{tcolorbox}[colframe=black, colback=white, arc=3mm, boxrule=1pt, width=\textwidth]
    \textbf{Tactical Dimension Prompt}  
    \tcblower
    \textbf{Task Definition:} 
    
    [Rationales $\mathcal{R}_{\mathcal{M}_i}$ generated by Strategic Dimension]  
    
    Given the high-level framework provided above, further decompose each macro-level module into its specific sub-functions or components. Each sub-function should be clearly defined with a corresponding description of its role. The generated sub-functions must align with the modules defined in the TotalFramework.
    
    \textbf{Formatting Requirements:}
    
    (1) The output should be structured in JSON format.  
    
    \textbf{Expected Output Format:}
    
    \begin{verbatim}
{
    "Modules": [
    {
        "Module": "[Module Name]",
        "SubFunctions": [
        {
            "Function": "[Sub-function Name]",
            "Responsibility": "[Brief description of its role]"
        },
        {
            "Function": "[Sub-function Name]",
            "Responsibility": "[Brief description of its role]"
        }
        ]
    }
    ]
}
    \end{verbatim}

    \medskip
    
    \textbf{Outputs:} 
\end{tcolorbox}

\clearpage
\subsection{Operational Dimension Prompt}
\begin{tcolorbox}[colframe=black, colback=white, arc=3mm, boxrule=1pt, width=0.95\textwidth]
    \textbf{Operational Dimension Prompt}  
    \tcblower
    \textbf{Task Definations:} 
    
    [Rationales $\mathcal{R}_{\mathcal{F}_j}$ generated by Tactical Dimension]  
    
    Please generate the Python code implementation for the \texttt{\{TaskName\}} sub-function in the \texttt{\{ModuleName\}} module, just the code.

    \textbf{Format Requirements:}
    
    (1) Ensure the code follows PEP 8 standards
    
    (2) Ensure the output only contend code, no other content!
    
    \textbf{Example Output:}
    \medskip
    
    \begin{lstlisting}[language=Python]
def AddTask(description):
    """
    Add a new task to the task list.

    Parameters:
    - description (str): Description of the task
    """
    if not description.strip():
        print("Error: Task description cannot be empty.")
        return

    tasks = load_tasks()
    task_id = len(tasks) + 1
    task = {
        'id': task_id,
        'description': description,
        'completed': False,
        'created_at': datetime.now().isoformat()
    }
    tasks.append(task)
    save_tasks(tasks)
    print(f"Task added: {task_id} - {description}")
\end{lstlisting}
    
    \textbf{Outputs:} 
\end{tcolorbox}

\clearpage
\section{Self-Rectification Prompts}
\label{sub:Self-Rectification Prompts}
\subsection{Strategic Dimension Verification Prompt}
\begin{tcolorbox}[colframe=black, colback=white, arc=3mm, boxrule=1pt, width=0.95\textwidth]

    \textbf{Strategic Dimension Verification Prompt}  
    
    \tcblower

    \textbf{Original Strategic Dimension Prompt:} 

    [Original Strategic Dimension Prompt]

    \textbf{Current Rationale:} 
    
    [Rationale $\mathcal{R}_{\mathcal{M}_i}$ generated by Strategic Dimension]
    
    \textbf{Task Definition:} 

    Evaluate the generated decomposition of a given task into Strategic-level modules for rationality and completeness. Base your judgment on \textbf{Original Strategic Dimension Prompt} and the provided \textbf{Current Rationale} using the following criteria:

    (1) \textbf{Clarity}: Is each module name clear and representative of its purpose?
    
    (2) \textbf{Completeness}: Do the modules cover all aspects of the task without missing significant components?
    
    (3) \textbf{Logical Structure}: Are the responsibilities assigned to each module logically coherent and appropriately categorized?
    
    (4) \textbf{Efficiency}: Does the structure optimize for ease of task execution and integration between modules?

    Provide your judgment as a real number between 0 and 1, where:
    
    0: The decomposition is completely unreasonable or irrelevant.
    
    1: The decomposition is perfectly reasonable and comprehensive.
    
    Provide only the score as a single float number (X.YY), where X is in [0, 1].
    \medskip
    
    \textbf{Outputs:} 
\end{tcolorbox}

\clearpage

\subsection{Tactical Dimension Verification}
\begin{tcolorbox}[colframe=black, colback=white, arc=3mm, boxrule=1pt, width=\textwidth]

    \textbf{Tactical Dimension Verification}  

    \tcblower

    \textbf{Original Tactical Dimension Prompt} 

    [Original Tactical Dimension Prompt]

    \textbf{Current Rationale:} 
    
    [ModuleName and Responsibility]

    \textbf{Tactical Decomposition}

    \textbf{Task Definition:}
    
    Evalute the generated decomposition of a given task into Tactical-level modules for rationality and completeness.\textbf{Current Rationale} using the following criteria:

(1) \textbf{High}: If the \textbf{Current Output} fully meets the expectations outlined in the PreviousLayerPrompt, assign a high score ($\ge$0.8).

(2) \textbf{Moderate}: If the \textbf{Current Output} partially meets the expectations (e.g., some components are vague or incomplete), assign a moderate score (between 0.4 and 0.7) with explanations for the deductions.

(3) \textbf{Low}: If the \textbf{Current Output} fails to meet the core requirements (e.g., lacks decomposition, missing responsibilities, or irrelevant content), assign a low score ($\le$0.3) with a brief explanation of the major shortcomings.

    Provide your judgment as a real number between 0 and 1, where:
    
    0: The decomposition is completely unreasonable or irrelevant.
    
    1: The decomposition is perfectly reasonable and comprehensive.
    
    Provide only the score as a single float number (X.YY), where X is in [0, 1].
    \medskip
    
    \textbf{Outputs:} 
\end{tcolorbox}

\clearpage

\subsection{Operational Dimension Verification}
\begin{tcolorbox}[colframe=black, colback=white, arc=3mm, boxrule=1pt, width=\textwidth]

    \textbf{Operational Dimension Verification}  

    \tcblower

    \textbf{Original Operational Dimension Prompt} 

    [Original Operational Dimension Prompt]

    \textbf{Current Rationale:} 
    
    [Function Definition and Responsibility]

    \textbf{Task Definition:}
    
    Please evaluate the generated code based on the provided function definitions and their responsibilities. The generated code must adhere to the following criteria:

(1) \textbf{Functionality Alignment:} Ensure that the \textbf{} accurately implements the described functionalities of each function. Penalize any deviation from the provided \textbf{responsibilities}.

(2) \textbf{Code Readability:} Assess whether the \textbf{code} is clear and well-organized, with proper variable naming, indentation, and documentation (if applicable).

(3) \textbf{Error Handling:} Verify that the code includes appropriate error-handling mechanisms where necessary. Penalize missing or insufficient handling for potential edge cases.

(4) \textbf{Modularity:} Evaluate the separation of concerns in the \textbf{generated code}. \textbf{Functions} should be self-contained and focus solely on their defined \textbf{responsibilities}, avoiding unnecessary coupling.

(5) \textbf{Efficiency:} Consider the computational efficiency of the \textbf{generated code}. Penalize unnecessary complexity or suboptimal logic.

(6) \textbf{Compliance with Standards:} Verify that the code adheres to the appropriate coding standards or style guides, including syntax correctness and consistent conventions.

    Provide your judgment as a real number between 0 and 1, where:
    
    0: The decomposition is completely unreasonable or irrelevant.
    
    1: The decomposition is perfectly reasonable and comprehensive.
    
    Provide only the score as a single float number (X.YY), where X is in [0, 1].
    \medskip
    
    \textbf{Outputs:} 
\end{tcolorbox}

\subsection{Rectification Prompt}
\begin{tcolorbox}[colframe=black, colback=white, arc=3mm, boxrule=1pt, width=0.95\textwidth]
    \textbf{Prompt}  
    \tcblower
    
    We have provided the previous Prompt and the output generated by the large model (Output). 
    Since the Output does not meet expectations, we need you to generate a new, more reasonable and higher-quality Output. 
    The new Output should better meet the requirements and address the issues in the previous Output.

    \textbf{Provided Prompt:} 
    
    [PreviousPrompt $\mathcal{P}_{\text{previous}}$]
    
    \textbf{Previous Output:} 
    
    [PreviousOutput $\mathcal{O}_{\text{previous}}$]
    \medskip
    
    \textbf{NewOutputs:} 
\end{tcolorbox}

\clearpage

\section{Case Study}
\label{sec:case study}
\begin{tcolorbox}[colframe=black, colback=white, arc=3mm, boxrule=1pt, width=0.95\textwidth]
    \textbf{Case Study}  
    
    \tcblower

    \textbf{Provided Prompt:} 
    
    \textbf{Task Definition:} 
    
    You are tasked with building a Python command-line application that simulates a “Project Management System.” The system should allow users to manage multiple projects and tasks, with features such as task allocation, project status updates, and user permissions. It should support multiple users, manage team members, and include functionality like multi-level task status tracking and permission management.
    
    \textbf{Key Features:} 
    
    Create Project: Users should be able to create new projects by providing project details such as name, description, deadline, priority, etc. Each project can contain multiple tasks.
Delete Project: Users should be able to delete a project using its unique project ID. Deleting a project should also remove all tasks associated with it.
Update Project Status: Projects can have different statuses like “In Progress,” “Completed,” or “Paused.” Users should be able to update the project’s status.
View Project Details: Users can view all projects in a summary list, or view detailed information of a specific project (including the task list, task status, and progress).
Assign Tasks: Each project can have multiple tasks, and tasks can be assigned to specific team members. Each task includes a title, description, assignee, status (e.g., “To Do,” “In Progress,” “Completed”), priority, and deadline.
Update Task Status: Task status can be updated. The available statuses are “To Do,” “In Progress,” and “Completed.” Admins and assigned team members should be able to change the task status.
Task Priority and Deadline: Tasks should have a priority level (e.g., “High,” “Medium,” “Low”) and a due date. Users should be able to filter tasks by priority or due date.
User Permissions: There are two roles—Admin and Regular User. Admins can create projects, assign tasks, delete projects, while regular users can only view projects and update the status of tasks they are assigned to.
Team Member Management: Each project can have multiple team members. Team members can view the project’s progress and update the status of the tasks they are responsible for.
Data Persistence: The project and task data should be saved to a local file (such as JSON or CSV) and loaded when the application starts. Any changes should be saved when the application closes.
Search Functionality: Users should be able to search projects by project name, tasks by title, or filter tasks by assignee.
Project Progress Report: The system should show a progress report for a project, including the count of completed tasks, pending tasks, and total tasks.

    \textbf{Technical Specifications:}

    (1) Use Python language to generate code.
    
    \medskip
    
    \textbf{Outputs:} 

    \textbf{Authorization and Access Control/denyUnauthorizedAccess.py}
    \begin{lstlisting}[language=Python]
    def deny_unauthorized_access(action: str, user_role: str, required_role: str):
    """
    Denies unauthorized access to specific actions based on user's role.
    Parameters:
    - action (str): The action to be performed (e.g., 'createProject', 'updateTask')
    - user_role (str): Current user's role (e.g., 'Admin', 'Regular User')
    """
    if user_role != required_role:
        raise PermissionError(f"Unauthorized access denied. Action '{action}' requires {required_role} role.")
    \end{lstlisting}
\end{tcolorbox}

\clearpage
\begin{tcolorbox}[colframe=black, colback=white, arc=3mm, boxrule=1pt, width=0.95\textwidth]
    
    \textbf{Authorization and Access Control/checkPermission.py}
    \begin{lstlisting}[language=Python]
    def check_permission(user_role, action, resource_type, resource_id=None):
    """
    Check if the user with the given role has permission to perform the specified action on the resource.

    Parameters:
    - user_role (str): The role of the user (e.g., 'Admin', 'Regular User')
    - action (str): The action to be performed (e.g., 'createProject', 'updateTask')
    - resource_type (str): The type of resource (e.g., 'project', 'task')
    - resource_id (int, optional): The ID of the resource, if applicable

    Returns:
    - bool: True if the user has permission, False otherwise
    """
    # Define a dictionary mapping roles to allowed actions for each resource type
    # This is a simplified example, in a real-world scenario, you would likely fetch these permissions from a database or config file
    allowed_permissions = {
        'project': {
            'Admin': ['createProject', 'updateProject', 'deleteProject', 'listProjects'],
            'Regular User': ['viewProjectProgress']
        },
        'task': {
            'Admin': ['createTask', 'updateTask', 'assignTask', 'trackTaskProgress'],
            'Regular User': ['assignTask', 'viewTaskProgress']
        }
    }

    if user_role not in allowed_permissions[resource_type]:
        return False

    if action not in allowed_permissions[resource_type][user_role]:
        return False

    # If resource_id is required and not provided, deny access
    if resource_id is not None and resource_id < 1:
        return False

    return True
    \end{lstlisting}
\end{tcolorbox}

\clearpage

\begin{tcolorbox}[colframe=black, colback=white, arc=3mm, boxrule=1pt, width=0.95\textwidth]
    
    \textbf{Authorization and Access Control/ModuleMerged.py}
    \begin{lstlisting}[language=Python]
    # permissions.py
def check_permission(user_role, action, resource_type, resource_id=None):
    
    # Define a dictionary mapping roles to allowed actions for each resource type
    # In a real-world scenario, fetch these permissions from a database or config file
    allowed_permissions = {
        'project': {
            'Admin': ['createProject', 'updateProject', 'deleteProject', 'listProjects'],
            'Regular User': ['viewProjectProgress']
        },
        'task': {
            'Admin': ['createTask', 'updateTask', 'assignTask', 'trackTaskProgress'],
            'Regular User': ['assignTask', 'viewTaskProgress']
        }
    }

    if user_role not in allowed_permissions[resource_type]:
        return False

    if action not in allowed_permissions[resource_type][user_role]:
        return False

    # If resource_id is required and not provided, deny access
    if resource_id is not None and resource_id < 1:
        return False

    return True


def deny_unauthorized_access(action: str, user_role: str, required_role: str):
    if user_role != required_role:
        raise PermissionError(f"Unauthorized access denied. Action '{action}' requires {required_role} role.")


# main.py
from permissions import check_permission, deny_unauthorized_access

def main():
    try:
        # Example usage
        user_role = "Admin"
        action = "createProject"
        resource_type = "project"

        # Check permission
        if not check_permission(user_role, action, resource_type):
            print("User does not have permission to perform the action.")
        else:
            print("User has permission to perform the action.")

        # Deny unauthorized access
        deny_unauthorized_access("createTask", user_role, "Admin")
    except PermissionError as e:
        print(e)

if __name__ == "__main__":
    main()
    \end{lstlisting}
\end{tcolorbox}

\clearpage

\begin{tcolorbox}[colframe=black, colback=white, arc=3mm, boxrule=1pt, width=0.95\textwidth]
    
    \textbf{Command-Line Interface (CLI)/handleInput.py}
    \begin{lstlisting}[language=Python]
import argparse

def handle_input(command: str, user_input: str):
    """
    Receive user input through CLI and interpret commands.

    Parameters:
    - command (str): The command requested by the user
    - user_input (str): Additional input required for specific commands
    """
    parser = argparse.ArgumentParser(description="User Command Interpreter")

    # Define subparsers for different commands
    subparsers = parser.add_subparsers(title="Commands", dest="command")

    # RegisterUser command
    register_parser = subparsers.add_parser("register", help="Register a new user")
    register_parser.add_argument("username", type=str, help="Username")
    register_parser.add_argument("password", type=str, help="Password")
    register_parser.add_argument("role", choices=["admin", "regular"], help="User role")

    # AuthenticateUser command
    authenticate_parser = subparsers.add_parser("authenticate", help="Authenticate user")
    authenticate_parser.add_argument("username", type=str, help="Username")
    authenticate_parser.add_argument("password", type=str, help="Password")

    # AddTask command
    add_task_parser = subparsers.add_parser("add_task", help="Add a new task")
    add_task_parser.add_argument("description", type=str, help="Task description")

    # Other commands can be added similarly...

    try:
        parsed_args = parser.parse_args(command.split())
        if parsed_args.command == "add_task":
            add_task(parsed_args.description)
        else:
            # Implement other command handling logic here...
            pass
    except argparse.ArgumentError as e:
        print(f"Invalid input: {e}")
    \end{lstlisting}
\end{tcolorbox}

\clearpage

\begin{tcolorbox}[colframe=black, colback=white, arc=3mm, boxrule=1pt, width=0.95\textwidth]
    
    \textbf{Command-Line Interface (CLI)/executeCommand.py}
    \begin{lstlisting}[language=Python]
import argparse
import datetime

def execute_command(command, args):

    # Command handlers (add more as needed)
    command_handlers = {
        'registerUser': handle_register_user,
        'authenticateUser': authenticate_user,
        'assignRole': assign_role,
        'updateUserDetails': update_user_details,
        'createProject': create_project,
        'updateProject': update_project,
        'deleteProject': delete_project,
        'listProjects': list_projects,
        'createTask': create_task,
        'updateTask': update_task,
        'assignTask': assign_task,
        'trackTaskProgress': track_task_progress,
        'saveData': save_data,
        'loadData': load_data,
        'backupAndSync': backup_and_sync,
        'handleInput': handle_input,
        'searchProjects': search_projects,
        'searchTasks': search_tasks,
        'generateReport': generate_report,
        'displayReport': display_report,
        'assignTeamMember': assign_team_member,
        'viewProjectProgress': view_project_progress,
        'updateTaskAssignment': update_task_assignment,
        'checkPermission': check_permission,
        'denyUnauthorizedAccess': deny_unauthorized_access,
    }

    if command in command_handlers:
        return command_handlers[command](args)
    else:
        raise NotImplementedError(f"Command '{command}' not recognized.")

# Implement each handler function here
def handle_register_user(args):
    # Implementation for registerUser
    pass

def handle_input(args):
    # Interpretation of user input and call executeCommand with appropriate command
    parsed_command = args.command
    parsed_args = parse_command_args(parsed_command, args.input)
    execute_command(parsed_command, parsed_args)

# Helper function to parse command and its arguments
def parse_command_args(command, user_input):
    # Example: using argparse for simple parsing
    parser = argparse.ArgumentParser()
    # Add parsers for each command
    subparsers = parser.add_subparsers(dest='command')
    register_user_parser = subparsers.add_parser('registerUser')
    register_user_parser.add_argument('username', ...)
    register_user_parser.add_argument('password', ...)
    return parser.parse_args(user_input.split())

\end{lstlisting}
\end{tcolorbox}

\end{document}